\documentclass[aip,rsi,amsmath,amssymb,reprint]{revtex4-1}%,twocolumn
\usepackage{graphicx}% Include figure files
\usepackage{url}
\usepackage{dcolumn}% Align table columns on decimal point
\usepackage{bm}% bold math
\usepackage{times}
\usepackage{tabularx}
\usepackage[colorlinks=true,allcolors=blue]{hyperref}
\begin{document}
	
	\preprint{AIP/123-QED}
	
	\title[]{Quantum dissipative systems beyond the standard harmonic model: features of linear absorption and dynamics}% Force line breaks with \\
	\author{Luke D. Smith}%Lines break automatically or can be forced with \\
	\email[]{cmlds@leeds.ac.uk}
	\affiliation{School of Chemistry, University of Leeds, Leeds LS2 9JT, United Kingdom%\\This line break forced with \textbackslash\textbackslash
	}%
	\author{Arend G. Dijkstra}%
	\email[]{A.G.Dijkstra@leeds.ac.uk}
	\affiliation{School of Chemistry, University of Leeds, Leeds LS2 9JT, United Kingdom%\\This line break forced with \textbackslash\textbackslash
	}%
	\affiliation{School of Physics and Astronomy, University of Leeds, Leeds LS2 9JT, United Kingdom%\\This line break forced with \textbackslash\textbackslash
	}%
	
\begin{abstract}
Current simulations of ultraviolet-visible absorption lineshapes, and dynamics of condensed phase systems, largely adopt a harmonic description to model vibrations. Often, this involves a model of displaced harmonic oscillators that have the same curvature. Although convenient, for many realistic molecular systems this approximation no longer suffices. We elucidate non-standard harmonic, and anharmonic effects, on linear absorption and dynamics using a stochastic Schr\"{o}dinger equation approach to account for the environment. Firstly, a harmonic oscillator model with ground and excited potentials that differ in curvature is utilised. Using this model, it is shown that curvature difference gives rise to an additional sub-structure in the vibronic progression of absorption spectra. This effect is explained, and subsequently quantified, via a derived expression for the Franck-Condon coefficients. Subsequently, anharmonic features in dissipative systems are studied, using a Morse potential, and parameters that correspond to the diatomic molecule $H_{2}$ for differing displacements and environment interaction. Lastly using a model potential, the population dynamics and absorption spectra for the stiff-stilbene photoswitch is presented and features are explained by a combination of curvature difference and anharmonicity in the form of potential energy barriers on the excited potential. 
\end{abstract}

\maketitle

\section{\label{sec:level1}Introduction}
The quantum dynamics of a system interacting with an environment is important in many fields of research. Prominent examples of this are found in excitonic energy transport in photosynthesis \cite{Panitchayangkoon2010,Scholes2010,Engel2007,Ishizaki2012, Caruso2009, Knee2017, Dijkstra2012, Duan2017, Wilkins2015, Baghbanzadeh2016} and the photoisomerisation event of molecular photoswitches \cite{Kumpulainen2017, Szymanski2013, Ikeda2019}, a key feature in the primary step of vision \cite{Palczewski2014,Palczewski2012, Johnson2015, Johnson2017, Mohseni2014, Hahn2000, Hahn2002, Farag2018}. The presence of the environment, which could be the solution in which a chemical reaction occurs or a protein, introduces the effects of relaxation and dephasing \cite{Breuer2007}. As a result, wavepacket dynamics along a potential energy surface are altered from the closed system evolution, as defined by the Schr\"{o}dinger equation. In addition to the interaction with the environment, the shape of the potential itself is also key in determining the quantum dynamics.

Commonly, theoretical approaches to open quantum systems approximate vibrational degrees of freedom of the environment with harmonic oscillators. This approximation results in evenly spaced energy levels, and represents gaussian fluctuations in the weak coupling regime \cite{Feynman2000}. Although the harmonic model is a common choice, molecular potentials are, in general, anharmonic and there are many examples that exhibit significant anharmonicity such as light-harvesting and photosynthesis \cite{Scholes2000, Meldaikis2013, Rancova2014}, photoswitches \cite{Kumpulainen2017}, and small molecules \cite{LeRoy2006, Foldi2002}. The feature of anharmonicity can also become pronounced when there is a large displacement between the ground and excited state potentials involving a large nuclear motion. In such cases, parts of the potential far from equilibrium may be explored, and the harmonic approximation is less likely to hold.

Treatment of anharmonic behaviour has been tackled by stochastic environments \cite{Packwood2011, Jansen2007}, molecular dynamics simulations \cite{Olbrich2011, Kwac2004, Garrett-Roe2008}, and by including anharmonicity in the system potential \cite{Anda2016, Anda2018, GalestianPour2017}. In this study we use anharmonic system potentials whilst also including an interacting environment via the stochastic Schr\"{o}dinger equation \cite{Breuer2007, Gardiner2004, Gardiner2009}. A review of other such open quantum system methods has been produced by Breuer \cite{Breuer2015}, and also by de Vega \cite{DeVega2017}. Alternatively, there exist \textit{ab initio} methods, such as the Multiconfigurational Ehrenfest (MCE) method \cite{Mahkov2017} that provides treatment of a large number of quantum nuclear degrees of freedom. In addition to this, there is the \textit{ab initio} multiple cloning (AIMC) method \cite{Green2019}, which is capable of simulating ultrafast excited state quantum dynamics following photo-absorption.

The roots of modelling anharmonicity can be found in the work of Osad'ko, and Skinner and Hsu \cite{Osadko2011,Skinner1986}. In addition, Tanimura used a treatment first via perturbation to harmonic potentials \cite{Okumura1997} and subsequently conducted studies with Morse potentials \cite{Tanimura1997}. Anharmonicity and its effects can manifest itself in numerous ways. The shape of the entire potential can be important, as in the case of the Morse potential and generally in the case of polynomial potentials \cite{Anda2016}. Additionally, displaced harmonic oscillators that have different curvatures are accredited with giving rise to non-standard spectral features \cite{Anda2018, Fidler2013}. Another feature is related to finer details of the potential, such as barriers that perturb the energy levels, and give rise to local minima which can trap the wavepacket. Realistic systems, in the condensed phase, can include an interplay of all these features in addition to the interaction with the environment.
\begin{figure}
	\includegraphics{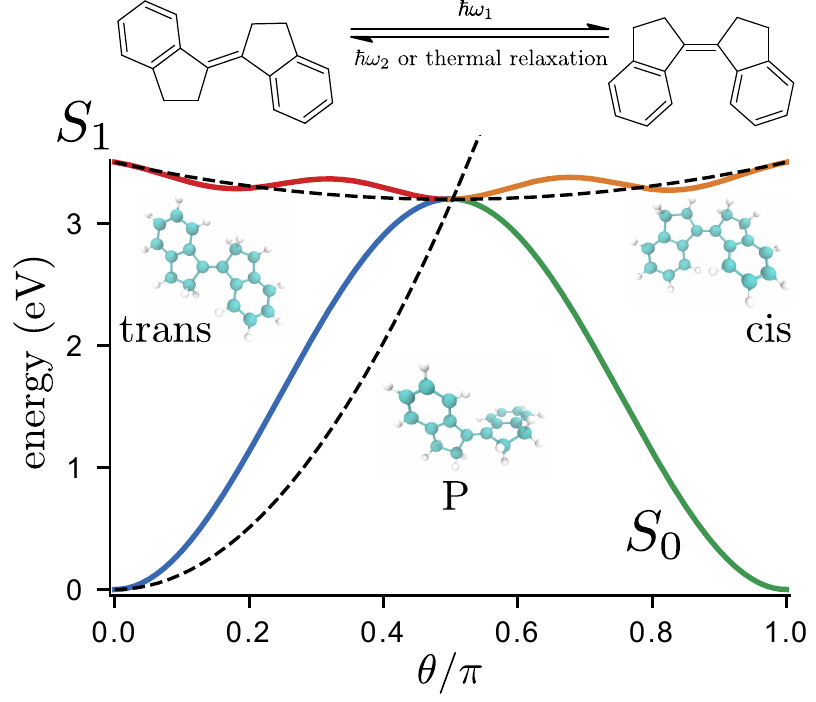}% Here is how to import EPS art
	\caption{\label{fig:PES} Model potential energy surface (PES) of stiff-stilbene in hexane as a function of the torsional coordinate $\theta$. The ground PES is represented by the curve $S_{0}$ and the excited PES by $S_{1}$. For comparison, a displaced harmonic oscillator model with differing curvatures, that approximates the stiff-stilbene PES, is represented by the black dashed lines. Light vertically excites the trans ground state at $\theta = 0$ to the excited state at approximately 3.5 eV. Subsequent rotation to $\theta = 0.5\pi$ takes it to the perpendicular conformation $P$, where there is a crossing point. Further rotation to $\theta = \pi$ leads to the cis conformation. Two important features are the presence of potential energy barriers on the excited state at $\theta = 0.3\pi$ and $\theta = 0.7\pi$, and the large difference in curvature of ground and excited potentials. }
\end{figure}

The presence of anharmonicity can have interesting effects on wavepacket dynamics. In the case of the Morse oscillator as displacement is increased, and anharmonic effects become more prevalent, a new phenomenon arises. The amplitude of oscillations of the expectation value of the position operator decreases to near zero and after a period of time revives to the near initial oscillation behaviour \cite{Foldi2002, Foldi2003, Benedict2003}. Such features can be observed in absorption and emission spectra, and also time-resolved nonlinear spectroscopies.

Various effects have been reported in absorption and emission spectra, such as mirror-symmetry breaking between absorption and fluorescence, and splitting of the zero phonon and one phonon peak \cite{Anda2016}. In 2D spectra the analogues of these effects have been studied as well as those not identified by linear spectra \cite{Anda2018}. Additionally, it has been shown that the ratio between selected cross peaks provides a measure of vibrational anharmonicity and other experimental indicators are possible \cite{GalestianPour2017}. 

It is known that spectral features may be broadened by the presence of an environment. A question remains as to how anharmonic and dissipative effects interplay, and the impact on the well known displaced harmonic oscillator model relations for absorption spectra.  Systems in which this might be particularly important include photoswitches, where there can be large displacements, and many of the stated features in the potential. These effects can be crucial in identifying spectral observables, wavepacket dynamics, and quantum yields.

In this paper we study the effects of anharmonicity, in the presence of an environment, on linear absorption spectra and wavepacket dynamics. We begin in Sec.\,\ref{sec:level2a} by introducing the photoexcitation model. Following this, treatment of the environment using the stochastic Schr\"{o}dinger equation is discussed in Sec.\,\ref{sec:level2b}. The theory of linear absorption is then described in Sec.\,\ref{sec:level2c}. In Sec.\,\ref{sec:level3a} we present the results of the harmonic differing curvature model, the resulting sub-structure in the vibronic progression, and Franck Condon coefficients that quantify this. Subsequently, in Sec.\,\ref{sec:level3b} we present the results for the Morse oscillator, spectral relations that fail due to dissipation and broadening, and the environment effects that are due to coupling to the anharmonic system. In Sec.\,\ref{sec:level3c} we present population dynamics and absorption spectra, for a model potential energy surface of the stiff-stilbene photoswitch, using the results of the previous sections to explain spectral features. The concluding remarks are then given in Sec.\,\ref{sec:level4}.
\section{\label{sec:level2}Theory}
\subsection{\label{sec:level2a}Model}
Throughout the paper we make use of electronic two-level systems, interacting with a bath, that depend on a single coordinate. The total Hamiltonian is given by \cite{Breuer2007}
\begin{equation}
H = H_{S} + H_{B} + H_{I}, \label{eq.total_hamiltonian}
\end{equation}
where $H_{B}$ represents the heat bath, $H_{I}$  is the interaction between system and bath, and the two-level system is represented by 
\begin{equation}
H_{S} =  H_{g} \vert g\rangle\langle g \vert + H_{e} \vert e\rangle \langle e \vert + J(\vert g\rangle \langle e \vert + \vert e \rangle \langle g \vert),
\end{equation}
where $\vert g \rangle$ and $\vert e \rangle$ represent ground and excited states respectively and
\begin{align}
H_{g} =& - \frac{\hbar^{2}}{2m} \frac{\partial^{2}}{\partial x^{2}} + S_{0}(x) \\
H_{e} =& E_{1} - \frac{\hbar^{2}}{2m} \frac{\partial^{2}}{\partial x^{2}} + S_{1}(x).
\end{align}
The coordinate of interest is represented by $x$, the momentum is given in terms of this coordinate and the mass $m$, $S_{0}(x)$ and $S_{1}(x)$ represent the ground and excited PESs, $J$ is the coupling between them, and $E_{1}$ provides the energy difference between the minima of the ground and excited state potentials. For simplicity, we assume that $J$ is independent of $x$. $H_{B}$ and $H_{I}$ describe the remaining environment degrees of freedom, and interaction with the system, which give rise to the effects of relaxation and dephasing. A common choice of PES, is the harmonic potential
\begin{equation}
S_{i}(x) = \frac{1}{2}m\omega_{i}^{2}(x-\Delta x_{i})^{2} + E_{i} \label{eq.harmonic_excited},
\end{equation}
where the harmonic frequency is given by $\omega_{i}$, and  $\Delta x_{i}$ represents the displacement from the ground state potential.
Also, a well known anharmonic PES is the Morse potential
\begin{equation}
V_{M}(x) = D_{e}(1-e^{-\beta(x-\Delta x)})^{2},
\end{equation}
where $D_{e}$ is the well depth defined relative to the dissociation energy, and $\beta$ is associated with the width. Figure \ref{fig:PES} shows the PES for stiff-stilbene in hexane which is used in Sec.\,\ref{sec:level3c} of the paper. This has been produced using TD-DFT results \cite{Improta2005}, and a schematic potential \cite{Quick2014, Kumpulainen2017}, to create a model potential that matches key features. The ground potential surface is given by,
\begin{equation}
S_{0}(\theta) = \frac{1}{2} (E_{P} - (\lambda_{g} - \mu_{g})) (1-\cos(2\theta)) + S_{01}(\theta) \label{eq.S0},
\end{equation}
where $E_{P}=3.2$ eV is the energy on the $S_{0}(\theta)$ potential at the perpendicular conformation, for which $\theta = 0.5$, and $\lambda_{g}$, $\mu_{g}$, and $S_{01}(\theta)$ are involved in the confining well. The full description is provided in Appendix \ref{sec:level5a}.
The excited potential surface is given by,
\begin{align}
S_{1}(\theta) =& E_{T} + \eta_{e}(\cos(6\theta)-1) \nonumber \\
&+ S_{11}(\theta) + S_{12}(\theta) + S_{13}(\theta), \label{eq.S1}
\end{align}
where $E_{T}$ is the energy on the $S_{1}(\theta)$ surface at the trans conformation $\theta=0$, and the second term defines three wells and two barriers in the region $0 \leq \theta \leq \pi$ with a height of $2\eta_{e}$. $S_{11}(\theta)$ defines a confining well, $S_{12}(\theta)$ allows further control over the well depth in the region $0\leq \theta \leq \pi$, and $S_{13}(\theta)$ allows control over the relative heights of the barriers. Using this model the barrier heights are chosen to be $E_{B1} = 0.0806$ eV and $E_{B2} = 0.105$ eV for the trans and cis barriers respectively \cite{Quick2014}.  Further detail on these terms is provided in Appendix \ref{sec:level5a}.  
Thus far the given description in Eq.\,(\ref{eq.total_hamiltonian}) is general, details about the bath have not been specified and shall be discussed in the next section.  
\subsection{\label{sec:level2b}Stochastic Schr\"{o}dinger equation}
In the previous section a description of the two-level system with coupling was given, along with examples of system Hamiltonians used. We now provide the description of the bath and how it is incorporated via the stochastic Schr\"{o}dinger equation (SSE). It is common to model the environment as a heat bath consisting of harmonic oscillators such that \cite{Gardiner2009,Gardiner2004} 
\begin{equation}
H_{B} = \hbar \int_{-\infty}^{\infty} d \omega\,\omega b^{\dagger}(\omega)b(\omega),
\end{equation}
and
\begin{equation}
H_{I} = i\hbar\int_{-\infty}^{\infty}d\omega\,\kappa(\omega)[b^{\dagger}(\omega)L - L^{\dagger}b(\omega)],
\end{equation}
where $b(\omega)$ are boson annihilation operators for the bath that have the relation
\begin{equation}
\big[b(\omega), b^{\dagger}(\omega)\big] = \delta(\omega - \omega^{\prime}),
\end{equation}
where $L$ is a system operator, and $\kappa(\omega)$ represents the strength of the coupling of the bath modes to the system. In this formalism the rotating wave approximation has been made and the approximation that the range of $\omega$ in the integrals is $(-\infty, \infty)$. Additionally, an approximation commonly called the first Markov approximation is often made in which the coupling constant is assumed independent of the frequency such that
\begin{equation}
\kappa(\omega) = \sqrt{\frac{\gamma}{2\pi}}.
\end{equation}
In the low temperature regime, and with the aforementioned approximations, a quantum white noise formalism is obtained and a stochastic Schr\"{o}dinger equation may be defined as \cite{Breuer2007,Gardiner2004}
\begin{equation}
d\vert\psi(t)\rangle = D_{1}[\vert\psi(t)\rangle]dt + D_{2}[\vert\psi(t)\rangle]dW(t) \label{eq.qsd},
\end{equation}
in which $dW(t)$ is a Wiener process, and $D_{1}$ is called the drift term, given by 
\begin{align}
D_{1}[\vert\psi(t)\rangle] =& -\frac{i}{\hbar}H_{S}\vert\psi(t)\rangle \nonumber \\ &+\frac{\gamma}{2}\Big(\langle L + L^{\dagger}\rangle_{\psi}L \nonumber \\
&- L^{\dagger}L - \frac{1}{4}\langle L + L^{\dagger}\rangle^{2}_{\psi}\Big)\vert\psi(t)\rangle,
\end{align}
where $L$ and $L^{\dagger}$ are system operators called Lindblad (or jump) operators, $\gamma$ quantifies the strength of coupling to the bath, and $\langle L + L^{\dagger} \rangle_{\psi}$ is concise notation for $\langle \psi(t) \vert L + L^{\dagger} \vert \psi(t) \rangle$. $D_{2}$ is the diffusion term, which is given by
\begin{equation}
D_{2}[\vert\psi(t)\rangle] = \sqrt{\gamma}\Big(L - \frac{1}{2} \langle L + L^{\dagger}\rangle_{\psi}\Big)\vert\psi(t)\rangle.
\end{equation}
The Wiener process must represent independent Gaussian random variables, with zero mean, and a variance of $\Delta t$. This is satisfied if
\begin{equation}
\Delta W_{k} = \sqrt{\Delta t}\xi_{k},
\end{equation}
where $\Delta W_{k}$, and $\Delta t$, represent discretisations of the Wiener process and time respectively, and $\xi_{k}$ is a Gaussian distributed random variable that has a mean of zero and unit variance.  
The closed system evolution of $\vert\psi(t)\rangle$ is represented by the first term in $D_{1}[\vert\psi(t)\rangle]$, whereas the open system is incorporated through the additional terms and Lindblad operators. It should be noted that for simplicity the above equations include interaction defined by a single Lindblad operator only, though the extension to multiple interactions is possible.  With regard to the open system terms, the drift term represents the drift of the state vector, and the diffusion term represents the random fluctuations due to the interaction of the system with the environment \cite{IanPercival1998}. Equation \ref{eq.qsd} is also known as a quantum state diffusion equation and has a corresponding density matrix equation given by the Markovian Lindblad master equation (LME) \cite{Breuer2007} 
\begin{align}
\frac{d}{dt}\rho(t) = -& \frac{i}{\hbar}[H,\rho(t)] \nonumber \\ +& \gamma \Big(L\rho(t)L^{\dagger} - \frac{1}{2}L^{\dagger}L\rho(t) - \frac{1}{2}\rho(t)L^{\dagger}L\Big).
\end{align}
To formulate the SSE and the LME only the Hamiltonian and the Lindblad operators are required. The choice of the Lindblad operators is arbitrary, up to the requirement of being a system operator, and is chosen to represent desired phenomena \cite{Kosloff1997}. A common case is relaxation through resonant energy transfer between system and bath, for which the Lindblad operators are chosen to be the creation and annihilation operators of the systems manifold. One such example is the case of a damped quantum harmonic oscillator for which $L=a$ where $a$ represents the lowering ladder operator for the harmonic oscillator \cite{Gisin1992a,Breuer2015}.

To simulate the stochastic Schr\"{o}dinger equation an appropriate numerical method that can solve stochastic differential equations must be implemented. We make use of an extension of the fourth-order Runge-Kutta scheme \cite{Breuer2007}, and apply it to the SSE. This is performed, on the wavepacket dynamics, for many iterations of the stochastic process, and a Monte Carlo average is taken. 
\subsection{\label{sec:level2c} Linear absorption spectra}
A useful tool to experimentally study the simultaneous transitions between molecular electronic states and vibrations is optical spectra \cite{Mukamel1995, Valkunas2013}. Linear absorption spectra is viewed as an elementary experiment that allows the elucidation of vibronic structure. 
Utilising the model system of section \ref{sec:level2a} we assume that we have two potential energy surfaces $S_{0}(x)$ and $S_{1}(x)$, dependent on a coordinate $x$, and a displacement between them $\Delta x$. The object of interest for the calculation of absorption spectra is the dipole correlation function 
\begin{equation}
C_{\mu\mu}(t) = \langle \bar{\mu}(t)\bar{\mu}(0) \rangle,
\end{equation}
where the dipole operator is given by
\begin{equation}
\bar{\mu} = \vert g\rangle \mu_{ge} \langle e \vert + \vert e \rangle \mu_{eg} \langle g \vert,  
\end{equation}
for which the Condon approximation has been made. This approximation assumes that the dipole operator has no nuclear dependence and only acts on the electronic states. The implication is that electronic transitions occur without a change of nuclear coordinate and the shape of the wavepacket remains unchanged, this is commonly known as a vertical transition due to how it looks on a potential energy diagram \cite{Yuen-Zhou2014}.

An important component of the correlation function, is given by the dephasing function \cite{Schatz2002, Reimers1983}
\begin{equation}
F(t) = \langle \psi_{g}(t) \vert \psi_{e}(t) \rangle \label{eq.dephasing_function}, 
\end{equation}
where $\langle \psi_{g}(t) \vert$ is a wavepacket on the ground potential and $\vert \psi_{e}(t) \rangle$ is a wavepacket on the excited potential. This formula makes no assumption on the form of the potential and can be calculated if the nuclear dynamics on ground and excited state surfaces are known. The relation between the dipole correlation and dephasing functions is such that
\begin{equation}
C_{\mu\mu}(t) = \vert \mu_{eg} \vert^{2} e^{-i\omega_{eg}t}F(t),
\end{equation}
where $\hbar\omega_{eg}$, in the standard displaced harmonic oscillator model, is commonly defined as the energy difference between the minima of potentials.
The absorption lineshape is then simply the Fourier transform of the dipole correlation function \cite{Mukamel1995}
\begin{align}
\sigma_{abs}(\omega) =& \int_{-\infty}^{\infty}dt\,e^{i\omega t}C_{\mu\mu}(t) \nonumber\\
=&\vert \mu_{eg}\vert^{2} \int_{-\infty}^{\infty}dt\,e^{i(\omega - \omega_{eg})t}F(t) \label{eq.absorption}.
\end{align}
The spectrum produced by this has a progression of absorption peaks from the peak centred at $\omega_{eg}$, which represents the 0-0 transition and is often called the zero-phonon line (ZPL). The shape and intensity of the progression depends on the displacement $\Delta x$ of the PES. Specifically, the Huang-Rhys factor $D$ quantifies the coupling strength of the electronic states to the nuclear degree of freedom and is defined as
\begin{equation}
D=\Delta x^{2} \frac{m\omega_{g}}{2\hbar}.
\end{equation}
For the displaced harmonic oscillator the Franck-Condon principle dictates a well defined relationship between the Huang-Rhys parameter $D$ and vibronic transitions observed in linear absorption spectra. For $D = 0$ one peak is observed corresponding to the electronic energy gap $\omega_{eg}$. In the weak regime $D<1$, the dependence of the energy gap on the coordinate $x$ is low such that the ZPL is seen as the peak with the largest amplitude. Additionally, the amplitude of the vibronic progression falls off as $D^{n}/(n+1)$, where $n$ refers to the eigenstate number. Finally, in the strong regime $D>1$, the peak with the most amplitude corresponds to $n=D$ such that upon excitation from the ground state the average number of vibrational quanta is equal to the Huang-Rhys parameter.
\section{\label{sec:level3}Results}
\subsection{\label{sec:level3a} Harmonic oscillator with differing curvatures}
One of the notable features of the stiff-stilbene PES of Fig.\,\ref{fig:PES} is the large difference in curvature of ground and excited potentials. To isolate, and illustrate, the effect this may have on absorption spectra we utilise a model that has harmonic ground and excited potentials that differ in curvature. We will refer to this as the harmonic differing curvature model, and we shall refer to harmonic potentials with equal curvature as the standard harmonic model. For the harmonic differing curvature model, a previous study by Fidler and Engel \cite{Fidler2013} has found that the location of the absorption peak maximum, and the absorption width, are dependent on curvature difference. Specifically, for a shallower excited state, the location of the absorption peak maximum will slightly shift to lower frequencies, whilst the peak width will decrease.

In this study we lift the restriction of modest curvature difference and displacement of the excited state PES, which are not valid assumptions for some photoswitches such as stiff-stilbene, to show that new features arise in absorption spectra.
\begin{figure}[t]
	\includegraphics{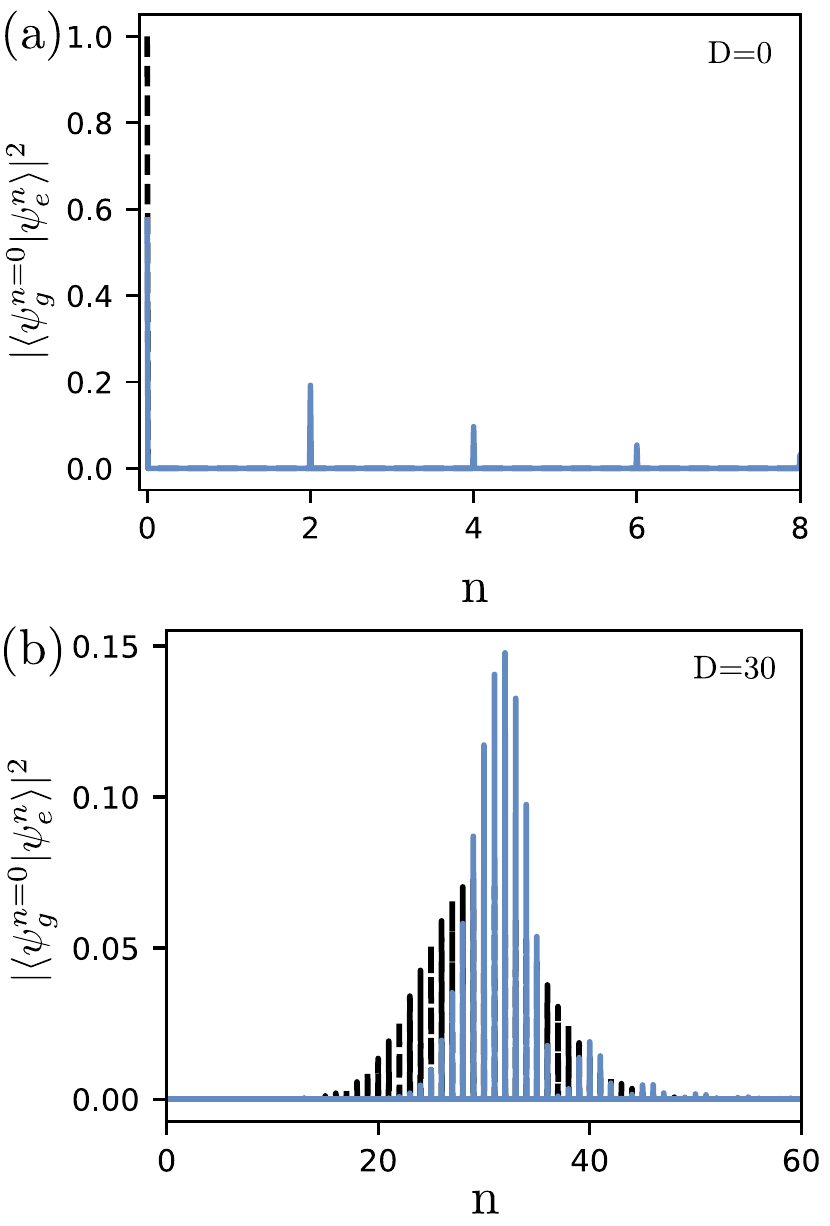}% Here is how to import EPS art
	\caption{Franck-Condon coefficients for the standard harmonic model (black dashed line) and differing curvature model (blue line) for (a) the case of no displacement and (b) displaced potentials. Notably, in the displaced and un-displaced cases, there is an additional vibronic sub-structure for the differing curvature model, not present in the standard harmonic model. Additionally, for the displaced case, there is a shift in the peak of the main progression to larger $n$ and the width of the progression decreases.} \label{fig.fcdc}
\end{figure}
\begin{figure}[b]
	\includegraphics{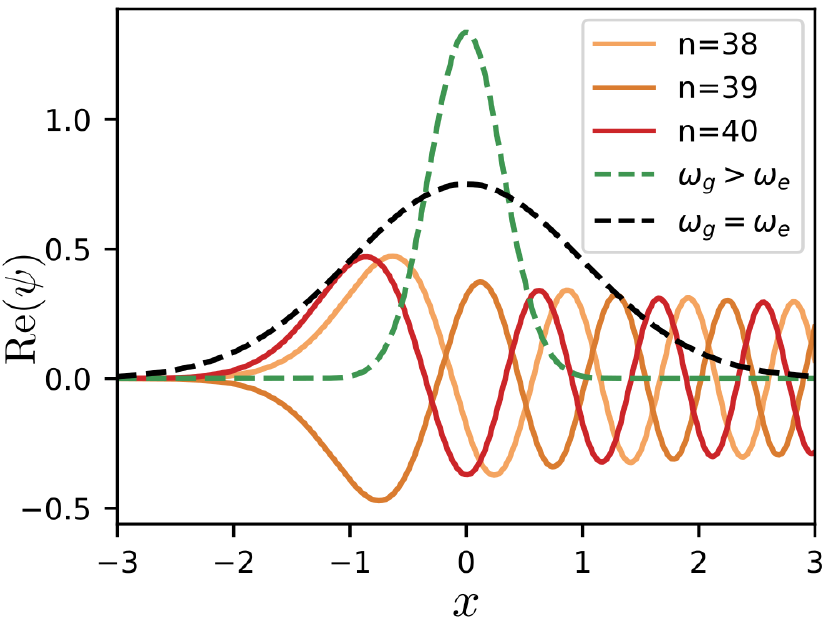}% Here is how to import EPS art
	\caption{The nuclear wavefunctions for the ground and excited potentials are represented by the dashed and solid lines respectively. For the differing curvature model (green dashed line) the overlap is sensitive to individual oscillations of wavefunctions on the excited potential. For $x=0$, if the excited state vibrational wavefunction $\vert \psi_{e}^{n} \rangle$ is close to a local minima or maxima, as in the case of $n=40$, the overlap is mainly constructive and leads to the large intensity parts of Fig.\,\ref{fig.fcdc}. The converse effect happens if at $x=0$, $\vert \psi_{e}^{n} \rangle$ is close to $0$, as in the case of $n=38$.} \label{fig.overlap}
\end{figure} 
To allow focus on the effect of curvature difference, spectral broadening effects of the environment, with this model, will not be included in this section, and instead presented in Sec.\,\ref{sec:level3c}. Under these circumstances the Franck-Condon principle explains the intensity of vibronic transitions that are shown in absorption spectra. This principle states that upon excitation, and associated electronic transition, a change from one vibrational energy level to another is dependent on the overlap of the nuclear wavefunctions, and more likely to occur if the overlap is significant. According to this principle, the amplitude of absorption peaks are given by the Frank-Condon coefficients
\begin{equation}
\vert \langle \psi_{g}^{n=0} \vert \psi_{e}^{n} \rangle \vert^{2}, 
\end{equation}
where $n$ represents the vibrational state of the nuclear wavepacket, $n=0$ represents the vibrational ground state wavepacket, and $g$ and $e$ represent the ground and excited electronic wavepackets respectively. This expression represents the overlap between the ground electronic state in the lowest vibrational state, and the excited electronic state in the $n^{th}$ vibrational state. The assumption that the ground state electronic wavepacket is in the lowest vibrational state holds for the low temperature regime, in particular it is valid for room temperature at $298$ K. In this case, the expression represents the intensities of peaks in low temperature absorption spectra.  
Franck-Condon coefficients for the standard displaced harmonic oscillator are well known and show a dependence on the Huang-Rhys parameter. For $D=0$ only one peak is expected corresponding to the coefficient
\begin{equation}
\vert \langle \psi_{g}^{n=0} \vert \psi_{e}^{n=0} \rangle \vert^{2} = 1.
\end{equation}
The peak intensity, corresponding to the respective Franck-Condon coefficient, is shown in Fig.\,\ref{fig.fcdc}a, where the standard harmonic case (black dashed line) is compared to the harmonic potentials with differing curvatures (blue line). Interestingly whilst in the standard harmonic case there is only one peak at $n=0$, in the harmonic with differing curvatures model a decaying progression of peaks is observed for even $n$, and the peak intensity is 0 for odd $n$. In the more general case that allows for non-zero displacement the Franck-Condon coefficient for the standard harmonic model is given by
\begin{equation}
\vert \langle \psi_{g}^{n=0} \vert \psi_{e}^{n} \rangle \vert^{2} = e^{-D} \frac{D^{n}}{n!} \label{eq.harmonic_FC},
\end{equation}
where the intensity is dependent on the Huang-Rhys parameter D. In Fig.\,\ref{fig.fcdc}b, for which $D=30$ and $\omega_{g} = \omega_{e} = 1$, this relation is seen to give rise to a Gaussian profile (black dashed line), centered at the Franck-Condon vertical transition. The largest intensity peak is at $n=30$ and thus $D$ can be associated with the mean number of vibrational quanta excited for $\vert \psi_{e} \rangle$. Comparing this to the harmonic differing curvature model, with $\omega_{g}=10\omega_{e}$ and $\omega_{e}=1$, there are three notable differences. Firstly, the width of the main vibronic progression decreases and the intensity of central peaks increases. Secondly, the largest intensity peak shifts from $n=30$ to $n>30$. Last but not least, there is the appearance of a number of smaller intensity vibronic progressions that occur at larger $n$ than the main vibronic progression, and these decay to zero as $n$ increases. Owing to the sub-structure nature of these progressions, we name this feature the sub-structure vibronic progression or s-progression for short.

Figure \ref{fig.overlap} allows insight into the appearance of the s-progression. Three excited state vibrational wavefunctions $\vert \psi_{e}^{n} \rangle$ are plotted corresponding to $n=38,39$ and $n=40$, this is compared to the ground state wavefunction $\vert \psi_{g}^{n=0}\rangle$ for the standard harmonic model $\omega_{g}=\omega_{e}$ and the differing curvature model $\omega_{g}>\omega_{e}$. The vibrational state numbers are chosen to correspond to where the first progression in the s-progression starts at approximately $n=38$ and to where it is at its maximum intensity at $n=40$. From this it can be seen that the narrow width of the ground wavefunction of the differing curvature model overlaps with close to only one of the oscillations of the excited state vibrational wavefunctions. As a result, the overlap is sensitive to whether, at $x=0$, the excited state vibrational wavefunction $\vert \psi_{e}^{n} \rangle$ is close to a local minima, maxima, or close to zero. For $n=38$, the peak of $\vert \psi_{g}^{n=0}\rangle$, located at $x=0$, is nearly aligned with a point where $\vert \psi_{e}^{n=38}  \rangle =0$, thus the overlap largely cancels and the Franck-Condon coefficient becomes close to zero. For $n=40$ the peak of the $\vert \psi_{g}^{n=0}\rangle$ is nearly aligned with a local minima of $\vert \psi_{e}^{n=40}\rangle$,  this results in a overlap that is large and thus this part of the s-progression is at its maximum intensity. In comparison, $\vert \psi_{g}^{n=0}\rangle$ in the standard harmonic model has a large width and does not pick out the fine structure of individual oscillations, therefore the effect is averaged out. 

The decay of the progression can also be understood through this figure, as $n$ increases the oscillations of $\vert \psi_{e}^{n}\rangle$, in the region of overlap, become closer together and equal in amplitude. For large $n$ the width of $\vert \psi_{g}^{n=0}\rangle$ in the differing curvature model no longer isolates individual oscillations and the effect decays to zero, as in the standard harmonic model. The shift of the main vibronic progression to larger $n$ can also be understood by Fig.\,\ref{fig.overlap}. The overlap occurs first at the edges of $\vert \psi_{e}^{n}\rangle$, and as the width of the differing curvature model is much less than in the standard harmonic model a larger $n$ is required before any overlap is achieved. The decrease in width of the main vibronic progression is also due to overlap with individual oscillations and the small width of $\vert \psi_{g}^{n=0}\rangle$, as the overlap increases and decreases more rapidly with increasing $n$.

Three features have thus far been identified, a shift of the main vibronic progression to larger $n$, a decrease in the width of the main progression, and the appearance of an s-progression. These have been explained with the help of Fig.\,\ref{fig.overlap} that shows the relative overlap. We now derive and present analytic expressions for the Franck-Condon coefficients that quantify these features and provide deeper insight into their appearance. We start with the case of $D=0$, where details of the derivation are contained in Appendix B. For the differing curvature model, the ground state wavefunction is given by
\begin{equation}
\vert \psi_{g}^{n=0}\rangle = N_{g} \exp\bigg(-\frac{1}{2}\alpha_{g}x^{2}\bigg) \label{eq.harmonic_wavefunction},
\end{equation}  
where
\begin{equation}
N_{g} = \Big(\frac{\alpha_{g}}{\pi}\Big)^{1/4}, \label{eq.norm_g}
\end{equation}
and
\begin{equation}
\alpha_{g} = \frac{m\omega_{g}}{\hbar}, 
\end{equation}
where, $\omega_{g}$ is the angular frequency of the  electronic ground state oscillator. The excited state wavefunction, in the $n^{th}$ vibrational state, is given by
\begin{equation}
\vert \psi_{e}^{n}\rangle = N_{n} H_{n}(\sqrt{\alpha_{e}}x)\exp\bigg(-\frac{1}{2} \alpha_{e} x^{2}\bigg),
\end{equation}
where 
\begin{equation}
N_{n} = \bigg(\frac{\sqrt{\alpha_{e}}}{2^{n}n!\sqrt{\pi}}\bigg)^{1/2}, \label{eq.norm_e}
\end{equation}
and
\begin{equation}
\alpha_{e} = \frac{m \omega_{e}}{\hbar}.
\end{equation}
The object of interest is the overlap integral between $\vert \psi_{g}^{n=0}\rangle$ and $\vert \psi_{e}^{n}\rangle$, which is given by
\begin{align}
\langle \psi_{g}^{n=0} \vert  \psi_{e}^{n}\rangle=& N_{e}N_{n} \int_{-\infty}^{\infty}dx\,H_{n}(\sqrt{\alpha_{e}}x)\exp\bigg(-\alpha x^{2}\bigg), \label{eq.unsolved_eq}
\end{align}
where
\begin{equation}
\alpha = \frac{\alpha_{e}+\alpha_{g}}{2},
\end{equation}
and $H_{n}(x)$ is the $n^{th}$ Hermite polynomial. Note that for odd $n$, $H_{n}(x)$ is an odd function. The product of this odd function with the even gaussian function, is odd. Taking the integral of an odd function over a symmetric region results in zero. Thus we have our first result that for odd $n$, and $D=0$ the Franck-Condon coefficient is zero, as demonstrated in Fig.\,\ref{fig.fcdc}a.

The derivation presented in Appendix B results in the expression
\begin{align}
\vert \langle \psi_{g}^{n=0} \vert  \psi_{e}^{n}\rangle \vert^{2}=&  \frac{n!}{2^{n}((\frac{n}{2})!)^{2}}\frac{\sqrt{\alpha_{e}\alpha_{g}}}{\alpha}\bigg(1-\frac{\alpha_{e}}{\alpha} \bigg)^{n}, \label{eq.FCDC D=0}
\end{align}
for the FC factors in the case of $D=0$, even $n$, and allows for the differing curvature model. 
Equation \ref{eq.FCDC D=0} has some noteworthy features. Firstly, for $\omega_{e}=\omega_{g}$ we also have $\alpha_{e} = \alpha$ and in this case the only non-zero value for the Franck-Condon coefficient is when $n=0$. Therefore, as expected, in the harmonic limit Eq.\,(\ref{eq.FCDC D=0}) gives $\vert \langle \psi_{g}^{n=0} \vert  \psi_{e}^{n=0}\rangle \vert^{2}=1$. In addition, this expression provides the peak intensity of the progression shown in Fig.\,\ref{fig.fcdc}a, and predicts a decay as $n$ increases. Furthermore, the progression will sustain for larger $n$, if the difference in curvature is increased. In the large curvature difference limit, the shape of the progression will be predominantly determined by
\begin{equation} 
\frac{n!}{2^{n}((\frac{n}{2})!)^{2}}\frac{\sqrt{\alpha_{e}\alpha_{g}}}{\alpha}.
\end{equation}
Following in the same manner we now derive and present an analytic expression for the more general case of when the differing curvature model is displaced. The process of the derivation is presented in Appendix B for the interested reader. An even more general expression for the Franck-Condon factors of the differing curvature model was derived by Chang \cite{Chang2005}, which allows for $n\geq0$ for $\vert\psi_{g}^{n} \rangle$. However, the derivation presented here diverges from that of Chang, implementing the solution found in the un-displaced model, connecting the two solutions. Additionally, the end expression obtained is in a form that allows for the interpretation of the observed features in Fig.\,\ref{fig.fcdc}, and provides insight into the appearance of the s-progression.
Firstly, the excited state is redefined as 
\begin{align}
\vert \psi_{e}^{n}\rangle = N_{n} H_{n}(\sqrt{\alpha_{e}}(x-d))\exp\bigg(-\frac{1}{2} \alpha_{e} (x-d)^{2}\bigg),
\end{align}
where $d$ corresponds to the displacement of the potential. Using this definition the overlap integral,  between $\vert \psi_{g}^{n=0}\rangle$ and $\vert \psi_{e}^{n}\rangle$, is given by
\begin{align}
\langle \psi_{g}^{n=0} \vert  \psi_{e}^{n}\rangle=& N_{e}N_{n} \int_{-\infty}^{\infty}dx\,H_{n}(\sqrt{\alpha_{e}}(x-d)) \nonumber \\
&\times \exp\bigg(-\frac{1}{2}(\alpha_{g}x^{2} +\alpha_{e}(x-d)^{2}\bigg). \label{eq.overlap_recast}
\end{align}
Following the derivation in Appendix B from Eq.\,(\ref{eq.y_recast}), the Franck-Condon coefficients for the differing curvature model, that admits displacement, is given by 
\begin{align}
\vert \langle \psi_{g}^{n=0} \vert  \psi_{e}^{n}\rangle \vert^{2}=& \frac{1}{2^{n}n!}\frac{\sqrt{\alpha_{e}\alpha_{g}}}{\alpha}e^{-A} \nonumber \\
&\times \bigg\vert n!\sum_{l=0}^{\left \lfloor{n/2}\right \rfloor }\frac{(-1)^{l}}{l!(n-2l)!}(2\beta)^{n-2l} \nonumber \\
&\times \bigg(1-\frac{\alpha_{e}}{\alpha}\bigg)^{l} \bigg\vert^{2}. \label{eq.FCDDC}
\end{align}
At this point it is illuminating to consider this equation in limits of interest. Firstly, in the standard displaced harmonic model limit, that is to say of equal curvature, $\alpha_{e}=\alpha$. In this case, the only term that survives in the summation is when $l=0$. Furthermore, in this limit $A=D$ the Huang-Rhys parameter, and $\beta = \sqrt{D/2}$. Therefore, making these substitutions we obtain the familiar formula, for harmonic FC coefficients, Eq.\,(\ref{eq.harmonic_FC}).

The second limit of interest is when the displacement is zero. In this case, the only term that survives the summation in Eq.\,(\ref{eq.FCDDC}), is when $l=n/2$. Substituting this value in and simplifying reproduces Eq.\,(\ref{eq.FCDC D=0}), the result of the first derivation. A final limit of interest is when the curvature is large, for which the shape of the progression is predominantly determined by 
\begin{align}
\vert \langle \psi_{g}^{n=0} \vert  \psi_{e}^{n}\rangle \vert^{2}=& \frac{1}{2^{n}n!}\frac{\sqrt{\alpha_{e}\alpha_{g}}}{\alpha}e^{-A} \nonumber \\
&\times \bigg\vert n!\sum_{l=0}^{\left \lfloor{n/2}\right \rfloor }\frac{(-1)^{l}}{l!(n-2l)!}(2\beta)^{n-2l}\bigg\vert^{2}.
\end{align} 
At this point the motivation for the form given can be found, as the summation in this equation is simply an explicit form of the Hermite polynomial
\begin{align}
H_{n}(b) = n!\sum_{l=0}^{\left \lfloor{n/2}\right \rfloor }\frac{(-1)^{l}}{l!(n-2l)!}(2\beta)^{n-2l}.
\end{align}
Therefore, substituting this expression gives
\begin{align}
\frac{1}{2^{n}n!}\frac{\sqrt{\alpha_{e}\alpha_{g}}}{\alpha}e^{-A}H_{n}(\beta)^{2},
\end{align}
for large curvature difference.

The expression of Eq.\,(\ref{eq.FCDDC}) provides the peak intensity of the progression shown in Fig.\,\ref{fig.fcdc}b. Furthermore, by comparing to the limiting cases of this model we see that the s-progression, in the displaced case, arises due to the terms of the summation. The summation itself is a modified form of an explicit expression for the Hermite polynomial in which, comparatively, latter terms of the summation contribute less. Thus, as $n$ increases the additional contribution of this summation becomes less important and, along with the other contributions in the equation, leads to the decaying feature of the s-progression. As in the zero displacement case, for a larger difference in curvature, the s-progression sustains for larger $n$. 
\subsection{\label{sec:level3b} Morse oscillator}
\begin{figure*}
	\includegraphics{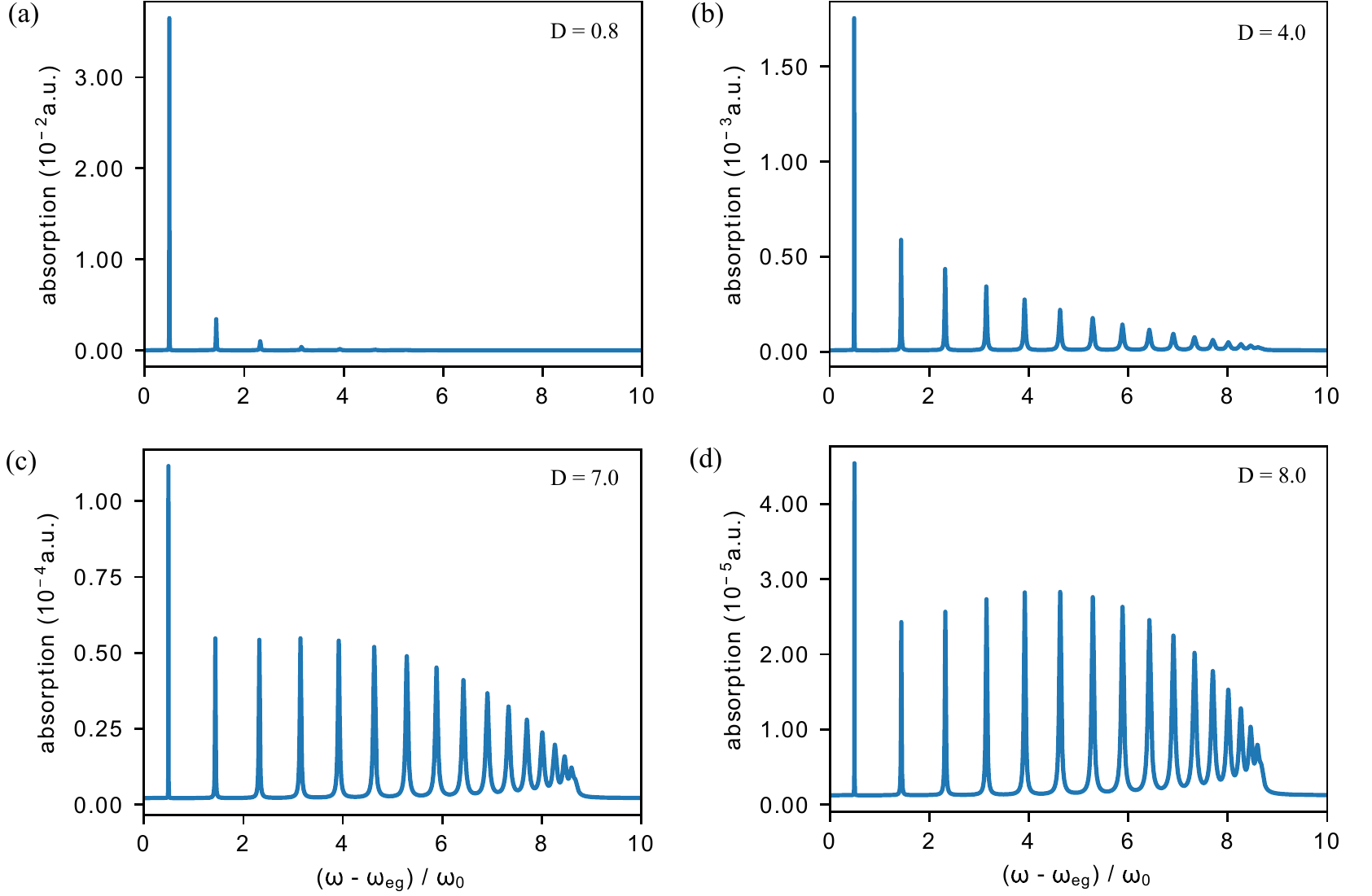}% Here is how to import EPS art
	\caption{\label{fig:spectra_D} A comparison of the linear absorption spectra of $H_{2}$ with differing values of Huang-Rhys parameter $D$. (a) Weak regime of $D=0.8$. (b) For $D=4.0$, a decaying vibronic progression is still observed. (c) For $D=7.0$, the behaviour changes as the amplitude of peaks becomes more uniform. (d) There is now a rise and fall of the amplitude of the vibronic progression. Peaks 5 and 6 have the largest amplitude, disregarding the ZPL.}
\end{figure*}
\begin{table}[b]
	\begin{center}
		\vspace{-4mm}
		\caption{Morse potential parameters of diatomic molecules}
		\label{tab:diatomic}
		\centering
		\setlength{\tabcolsep}{16pt}
		\begin{tabularx}{\columnwidth}{l c c c c c}
			\hline\hline\\
			Molecule & $D_{e}(E_{h})$ & $\omega_{x}(E_{h})$ & $\beta(a_{0})$\\ 
			\hline
			$B_{2}$ & 0.104 & 0.005 & 1.89\\
			$H_{2}$ & 0.174 & 0.020 & 1.95\\ 
			$O_{2}$ & 0.190 & 0.007 & 2.66\\
			$F_{2}$ & 0.064 & 0.004 & 2.75\\
			$N_{2}$ & 0.277 & 0.011 & 3.09\\
			\hline\hline
		\end{tabularx}
	\end{center}
\end{table}
To study the effects of anharmonicity on linear absorption spectra we first use a Morse potential with a harmonic frequency and dissociation energy aimed to represent the bond vibration of the $H_{2}$ molecule \cite{Micciarelli2018}. Parameters for $H_{2}$ and other molecules \cite{Hulburt1941, Huber1979}, for comparison, are contained in Table.\,\ref{tab:diatomic} in atomic units. The absorption spectra, simulated with weak dissipation, for differing values of $D$ is shown in Fig.\,\ref{fig:spectra_D}, where we define $\omega_{eg}=E_{1}/\hbar - \omega_{0}/2$, for which $\omega_{0}/2$ is subtracted to correct for the zero point energy of the ground state.  In the case of the Morse oscillator we find that many of the harmonic relations of Sec.\,\ref{sec:level2c} no longer hold. Firstly, the Franck-Condon factor for the vibrational ground state is larger and the first vibrational state is lower than in the Harmonic case. In the very low limits of $D$ we expect results to be close to harmonic as only parts of the potential close to the minimum are explored. As the Huang-Rhys parameter is increased towards $D=1$, as in Fig.\,\ref{fig:spectra_D}a, the Morse oscillator vibrational ground state still features a larger FC factor but the first and second states are less populated than in the harmonic case. Furthermore, higher lying states show increased FC factors.

These features can be explained due to the asymmetry of the Morse vibrational eigenfunctions that are skewed to the shallow side of the potential. In the weak regime, the higher states thus have greater overlap with a displaced wavefunction that is close to the ground state wavefunction, as is the case for $D<1$ \cite{Micciarelli2018}.

Another such phenomenon is that a diminishing intensity of the vibronic progression is still observed for $D>1$, as shown in Fig.\,\ref{fig:spectra_D}, contrary to harmonic observations. Also, in the strong regime the peak with the largest amplitude, disregarding the ZPL, occurs at a lower frequency than for harmonic spectra. These observations are due to the effects of asymmetric broadening of the spectra as each eigenstate relaxes at a different rate \cite{Foldi2003,Benedict2003}. In addition, the Morse distribution of the amplitude of FC factors is more uniform which makes the relative amplitudes of spectra more sensitive to asymmetric broadening. This also explains why the ZPL is observed as the largest peak as it features no asymmetric broadening. Analysis with no dissipation shows the peak with largest amplitude is the same as in the harmonic case.

\begin{figure}[t]
	\includegraphics{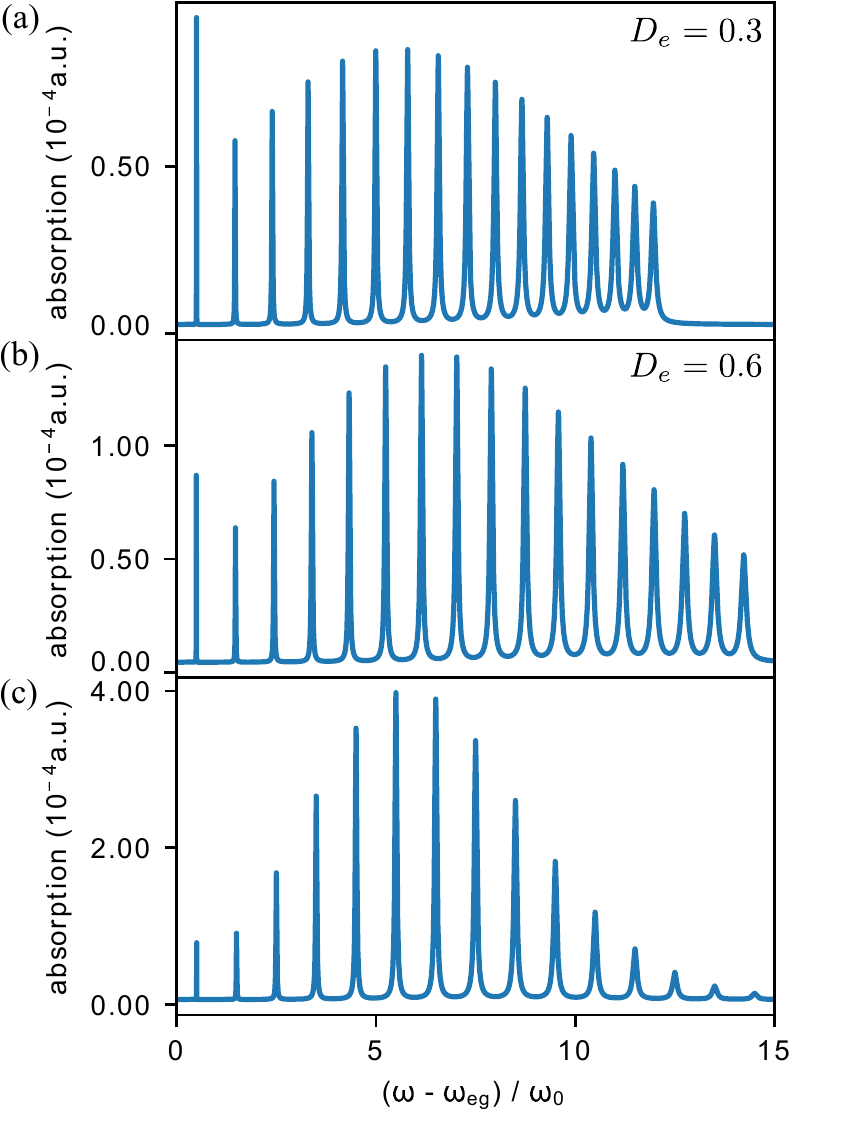}% Here is how to import EPS art
	\caption{\label{fig:spectra_harm} Linear absorption spectra for $D=7.0$ is shown for increasing dissociation energy $D_{e}$ whilst retaining other parameters of $H_{2}$. (a) The increase from $D_{e}=0.174$ to $D_{e} = 0.3$ introduces rise and fall behaviour of the amplitudes of the vibronic progression. (b) For $D_{e} = 0.6$, the central peak amplitudes increase such that relatively the ZPL no longer has the largest peak amplitude. (c) Harmonic absorption spectra. }
\end{figure}

To further exemplify these features we look at the harmonic limit in Fig.\,\ref{fig:spectra_harm} for $D=7.0$. To approach the harmonic limit the dissociation energy $D_{e}$ is increased whilst maintaining other parameters. As $D_{e}$ is increased the spectral lines become more evenly spaced which shifts the higher states to larger frequencies. However, as the results become more harmonic the distribution of amplitudes becomes less uniform, and more peaked around the center of the vibronic progression. For $D_{e}=0.6$ the peaks corresponding to energies of eigenstates at $n=6$ and $n=7$ have become the largest intensity peaks. The ZPL still remains greater in intensity than the harmonic case as do the wings of the vibronic progression due to a more uniform distribution of FC factors.

 As the harmonic limit is achieved the ZPL has decreased as have the wings of the spectra. The tallest peaks do not correspond to $n=6$ and $n=7$ due to asymmetric broadening effects but the overall shape of the vibronic progression is still close to a gaussian profile. These results suggest that the features of the Morse spectra are sensitive to the effects of dissipation, especially so in the case when $D>1.0$.

It is commonplace in the literature to use harmonic raising and lowering operators as the Lindbladians regardless of the actual raising and lowering operators of the system. The question is thus raised, as to whether using Morse raising and lowering operators \cite{Dong2002} creates significant difference on spectral features. 

\begin{figure}[b]
	\includegraphics{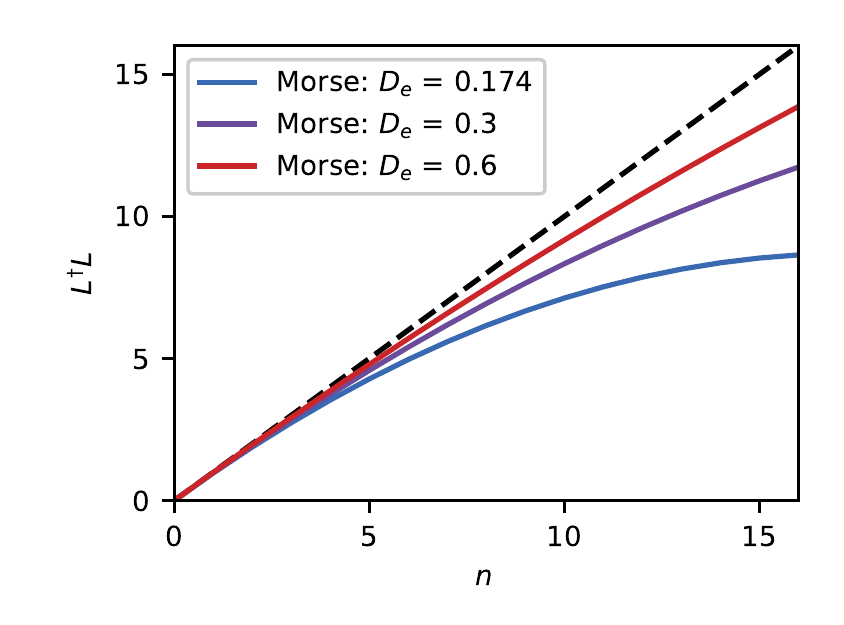}% Here is how to import EPS art
	\caption{\label{fig:lindblad_dissipation} The term $L^{\dagger}L$ which defines dissipation is shown against the number state for Morse raising and lowering operators. $H_{2}$ parameters are used and the dissociation energy is increased to observe the effect of increasing harmonicity. The harmonic case is represented by the black dashed line. }
\end{figure}

Using harmonic raising and lowering operators we have
\begin{align}
\Gamma L^{\dagger}L|n\rangle = \Gamma n |n\rangle, 
\end{align}
where $\Gamma$ controls the strength of dissipation, $L^{\dagger}$ and $L$ are Lindblad operators, and we note this choice of Lindbladian causes downwards transitions proportional to $n$. 

\begin{figure*}
	\includegraphics{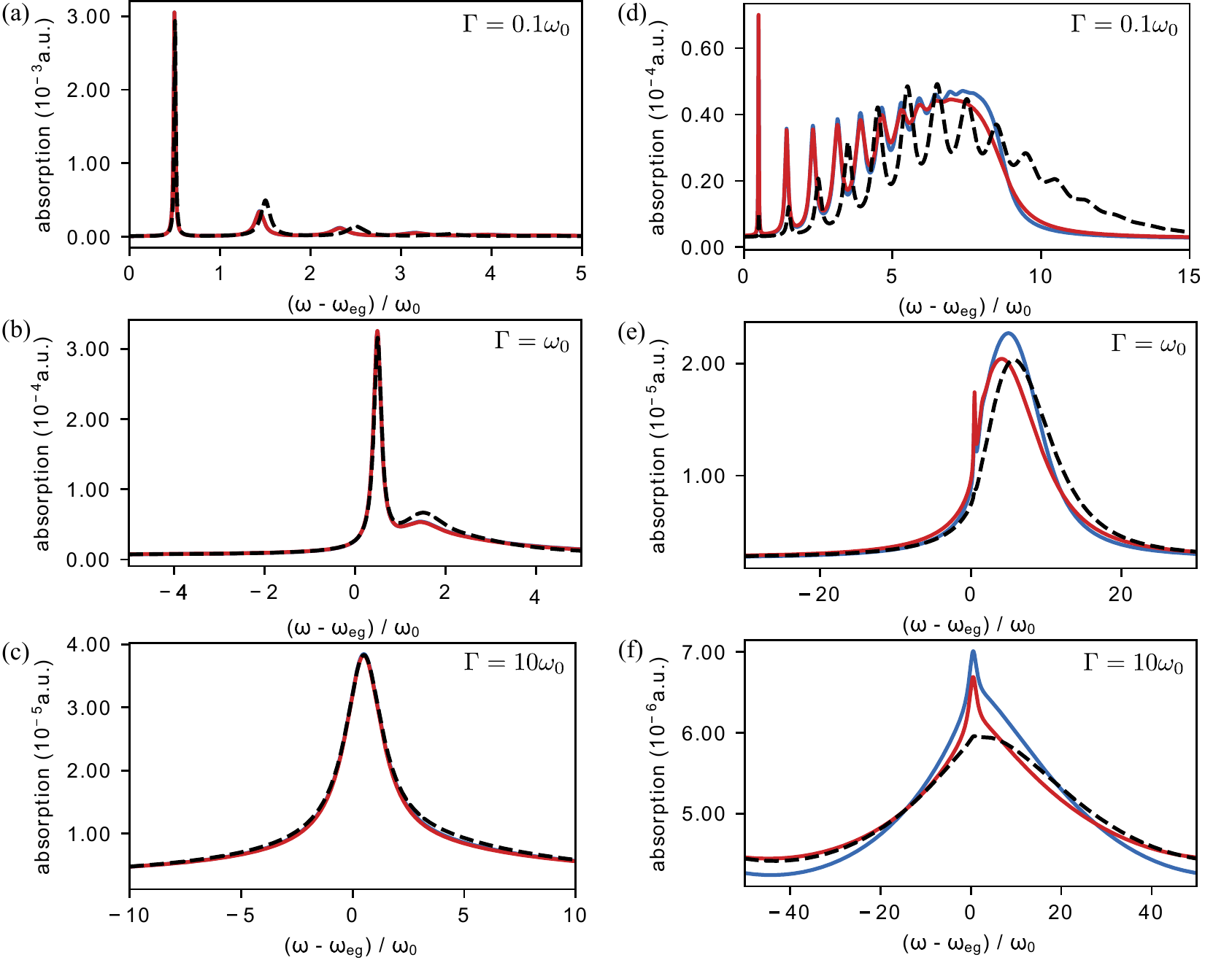}% Here is how to import EPS art
	\caption{\label{fig:spectra_diss_D_1} Linear absorption lineshape is shown for 3 increasing values of the dissipation parameter $\Gamma$. This is conducted for $D=1.0$ in (a), (b) and (c) and $D=7.0$ in (d),(e), and (f). The harmonic result (black dashed line), Morse result (blue line), and the Morse result using harmonic raising and lowering operators (red line) are shown to have little difference for the small Huang-Rhys parameter $D=1.0$. In comparison, for a large value of $D=7.0$ it is observed that the resulting lineshape is dependent on shape of the potential and the choice of Lindblad operator.}
\end{figure*}

For the Morse raising and lowering operators \cite{Dong2002} we have
\begin{align}
\Gamma L^{\dagger}L |n\rangle = \Gamma (n-\frac{n}{\nu})|n\rangle,
\end{align}

where $\nu$ is a measure of the systems harmonicity. This imposes that the higher lying states experience less dissipation than when using harmonic raising and lowering operators. The harmonic limit is achieved as $\nu\rightarrow\infty$. A comparison of these terms is shown in Fig.\,\ref{fig:lindblad_dissipation}. As $D_{e}$ is increased, and the system becomes more harmonic, the dissipation as defined by $L^{\dagger}L$ becomes linear in $n$. 

To study the effects of the different raising and lowering operators on absorption spectra we first choose a Huang-Rhys factor of $D=1.0$ , see Fig.\,\ref{fig:spectra_diss_D_1}(a),(b) and (c). This represents relatively low displacement. We study three choices of dissipation: weak $\Gamma = 0.1\omega_{0}$, medium $\Gamma = \omega_{0}$, and strong $\Gamma = 10\omega_{0}$. Results are shown in Fig.\,\ref{fig:spectra_diss_D_1} for harmonic spectra (black dashed line), Morse spectra with harmonic raising and lowering operators (red line), and Morse raising and lowering operators (blue line).

For the weakest dissipation a decaying vibronic progression is observed, with small differences between the Morse and harmonic spectra. The Morse ZPL is larger in amplitude and the ensuing peaks are less than for the harmonic spectra featuring a shift to lower frequencies due to closer vibrational states. For a medium level of dissipation we observe a ZPL which has larger amplitude for the Morse spectra and a phonon sideband with lower amplitude which is shifted to lower frequencies as compared to harmonic. For strong dissipation there is a broad Lorentzian lineshape which is shallower in the wings for the Morse spectra. A common feature among all dissipation levels is that, for $D=1.0$, there is almost no perceivable difference in using Morse raising and lowering operators. The small observed differences are due to the shape of the potential itself and the dissipation is well approximated via harmonic raising and lowering operators. It is also noteworthy that the harmonic case overestimates the Huang-Rhys factor if used to approximate Morse. 

We now turn our attention in Fig.\,\ref{fig:spectra_diss_D_1}(d),(e) and (f) to the large Huang-Rhys factor $D=7.0$ to demonstrate the effect of using Morse raising and lowering operators. The results for the Morse spectra have been normalised and much of population of the system, for such a large Huang-Rhys factor, is contained in the continuum of higher states. As we are interested in the anharmonic effects on the bound spectra we restrict ourselves to only the bound states of the system. In the low dissipation regime the ZPL for the Morse oscillator has a larger relative amplitude with respect to its vibronic progression. The peak intensity shows only a steady rise in comparison to the harmonic. For the low lying states $n=0,1,2$ there is minimal difference between using Morse raising and lowering operators, as demonstrated when using $D=1.0$. However, as $n$ increases differences become more apparent and we see that the peaks and troughs of the spectra, generated using Morse raising and lowering operators, are increased.

For medium dissipation the changes are further exemplified. The harmonic spectra features only a broad lineshape, whereas the Morse spectra still features a visible ZPL with large phonon sideband. The profile for the Morse lineshape is shifted to lower frequencies with respect to the harmonic frequency. The use of Morse raising and lowering operators has significant difference and we observe a lineshape with larger amplitude. The changes are more significant to higher frequencies in the spectra with rough agreement to harmonic raising and lowering operators at low frequencies. Notably the Morse spectra with harmonic raising and lowering operators has a very similar profile to harmonic spectra albeit shifted to lower frequencies and with a more defined ZPL. When the Morse raising and lowering operators are utilised there is less of a shift to lower frequencies and a different profile at higher frequencies such that it is at first steeper than the harmonic profile and then shallower. This overall effect causes the spectra to appear more symmetric about the broadened peak of the sideband. In the large dissipation regime the lineshape is very broad. This is due to asymmetric broadening that is not proportional to $n$, but instead to $n-n/\nu$. The harmonic spectra features one broad peak and hardly any definition to the ZPL. the Morse spectra features a more defined point to where the ZPL is with a protrusion of the broadening to higher frequencies. Lastly, the Morse raising and lowering operator spectra features a larger steep to shallow transition in its lineshape profile. 

\subsection{\label{sec:level3c}Stiff-stilbene}
\begin{figure*}
	\includegraphics{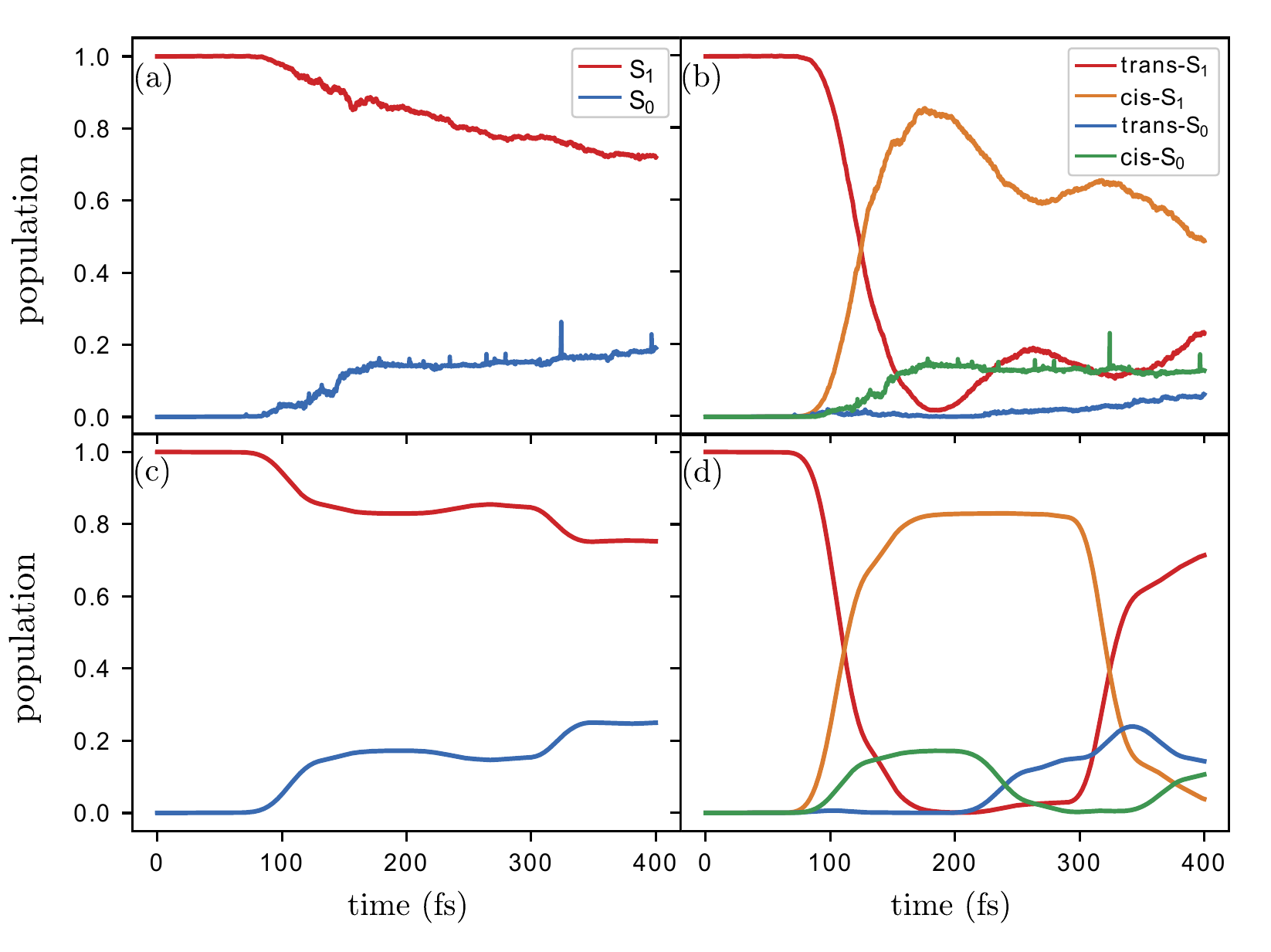}% Here is how to import EPS art
	\caption{Time evolution of the populations are shown for damped dynamics, in (a) and (b), and closed dynamics, in (c) and (d). The left column represents population on either the excited PES, $S_{1}$, or the ground PES, $S_{0}$. The right column displays the time evolution of finding the system in cis or trans conformations, and is colour coordinated to correspond to Fig.\,\ref{fig:PES}.   }\label{fig.stiff_populations}
\end{figure*}
Thus far, we have described how a large difference in curvature between the displaced ground and excited potentials can produce features in linear absorption spectra. Additionally, we have shown how anharmonicity can alter lineshape by changing the positioning and spacing of the vibronic progression, and how this is influenced by broadening caused by the environment.
In connection to this, the model PES of the stiff-stilbene photoswitch shown in Fig.\,\ref{fig:PES}, possesses both large curvature difference and anharmonicty. In this section the trans-cis population dynamics and absorption spectra of stiff-stilbene, using the developed model PES of Eq.\,(\ref{eq.S0}) and Eq.\,(\ref{eq.S1}), are interpreted using the results of the previous sections. Through this, features that are not captured by the standard harmonic model, of the dissipative dynamics and linear absorption, are explained.

Firstly, the closed system dynamics of the PES in Fig.\,\ref{fig:PES} are simulated to allow for comparison. To accomplish this, it is assumed that the system starts out in the ground state wavepacket $\vert\psi_{g}^{n=0} \rangle$, which is approximated as harmonic by fitting to the ground state potential, as in Eq.\,(\ref{eq.harmonic_wavefunction}), with a frequency $\omega_{g} = 0.085 $ fs$^{-1}$. Rotational analogues of motion are taken, such that the coordinate of interest depends on rotation about the double bond in stiff-stilbene $\theta$. In addition to this, the mass is replaced by the moment of inertia $I = 1001$ eVfs$^{-1}$, and the reduced planck constant in these units is $\hbar = 4.136/2\pi$ eVfs$^{2}$. A harmonic potential is fitted to the excited PES as in Fig.\,\ref{fig:PES}, with $\omega_{e} = 0.01571$ fs$^{-1}$. This frequency ensures the desired period of $400$ fs, and thus corresponds to a barrierless isomerisation time of $200$ fs \cite{Kovalenko2013,Quick2014}. In turn this allows the calculation of the moment of inertia stated,
\begin{equation}
I=\frac{2(S_{1}(\theta)-E_{1})}{\omega_{e}^{2}(\theta - \pi/2)^{2}},
\end{equation}
where the definition of the excited state potential Eq.\,(\ref{eq.harmonic_excited}) has been used with rotational analogues, and $E_{1} = 3.195$ eV. Following this, a point on the potential, for example $S_{1}(0) = 3.5$ eV \cite{Improta2005}, is substituted to find the value of the moment of inertia that is consistent with the expected period and model PES.
It is assumed that the wavepacket is vertically excited from the ground potential, thus maintaining the same shape at $t=0$ on the excited potential 
\begin{equation}
\vert \psi_{e}(t=0)\rangle = \vert \psi_{g}^{n=0} \rangle.
\end{equation}
Subsequently, the wavefunction is propagated forward in time using the time-dependent Schr\"{o}dinger equation
\begin{equation}
d\vert \psi_{e}(t)\rangle = -\frac{i}{\hbar}H_{S}\vert \psi_{e}(t)\rangle dt,
\end{equation}
which is a limiting case of the SSE Eq.\,(\ref{eq.qsd}) for $\gamma = 0$, and $H_{S}$ is the Hamiltonian for stiff-stilbene which has a coupling $ J=0.02$ eV, and potentials defined by $S_{0}(\theta)$ and $S_{1}(\theta)$ in Eq.\,(\ref{eq.S0}) and Eq.\,(\ref{eq.S1}) respectively. The simulation is carried out using a fourth-order Runge-Kutta scheme, for $\theta \in [-2\pi, 3\pi]$, with a spacing of $0.002\pi$. This ensures $500$ grid points in the range of interest $\theta \in [0, \pi]$, and a large enough grid to negate boundary effects. The wavepacket dynamics are computed for a time of $400$ fs, with a time step of $\Delta t = 0.001$ fs. The population dynamics of $S_{1}$ is calculated by
\begin{align}
P_{S_{1}}(t) = \int_{-\infty}^{\infty}d\theta\,\psi_{e}^{*}(\theta,t)\psi_{e}(\theta,t),
\end{align}
and similarly the population dynamics of $S_{0}$ is given by 
\begin{align}
P_{S_{0}}(t) = \int_{-\infty}^{\infty}d\theta\,\psi_{g}^{*}(\theta,t)\psi_{g}(\theta,t).
\end{align}
To calculate the trans and cis populations, for $S_{0}$ and $S_{1}$, the limits of integration are restricted to $\theta \in [0,\pi/2]$ and $\theta \in [\pi/2,\pi]$ respectively. The results of the closed dynamics are shown in Fig.\,\ref{fig.stiff_populations}c and Fig.\,\ref{fig.stiff_populations}d, which is colour coordinated to match Fig.\,\ref{fig:PES},  and can be interpreted as follows. The wavepacket initially starts out in the trans-$S_{1}$ conformation and moves along the potential, at approximately $100$ fs it reaches the perpendicular conformation. At this point the population transfers to cis-$S_{1}$ and also, due to the crossing, to the desired photoswitched state cis-$S_{0}$. Following the cis-$S_{1}$ population, it takes a further $200$ fs to reach the perpendicular conformation again. This occurs at $300$ fs, upon which population is transferred from cis-$S_{1}$ to the trans-$S_{0}$ state and trans-$S_{1}$. In addition to this, the population that transferred to cis-$S_{0}$ at $100$ fs has a faster period of oscillation, and at approximately $250$ fs the wavepacket on the ground potential reaches the crossing point, and the population transfers from cis-$S_{0}$ to trans-$S_{0}$.

To simulate the damped dynamics, the SSE of Eq.\,(\ref{eq.qsd}) is used. As in the closed case, the system is assumed to start in the ground state wavepacket $\vert\psi_{g}^{n=0} \rangle$, and is then vertically excited to the excited PES. There are some more parameters and operators for the damped case which must be first specified before the dynamics are simulated. These are the dissipation parameter, which is chosen to be $\gamma = 0.2\omega_{e}$, to ensure appropriate broadening in absorption spectra and significant population trapping in the cis-$S_{1}$ state at $400$ fs \cite{Quick2014}. In addition to this, the Lindblad operator for evolution on $S_{1}$ is chosen as the lowering operator of the harmonic fit to $S_{1}$. This results in damping towards the minima of $S_{1}$, corresponding to the perpendicular conformation. For $S_{0}$ Lindblad operators corresponding to the lowering operator of a harmonic potential fit at $\theta=0$, and a fit at $\theta=\pi$, are used to allow for damping towards the minima of the potential corresponding to the trans and cis states respectively.

The results of the damped dynamics are shown in Fig.\,\ref{fig.stiff_populations}a and Fig.\,\ref{fig.stiff_populations}b. Where the simulation implements an adaptation of the fourth-order Runge-Kutta scheme to SSEs \cite{Breuer2007}. The overall population dynamics on $S_{0}$ and $S_{1}$ behave in a similar manner to the closed evolution, specifically in the sense that at $400$ fs the populations are in close agreement. However, there is a difference in the population transfer such that it is more gradual in the damped case, whereas occurs in steps in the closed case, an explanation for this will be provided through consideration of the trans and cis populations. In Fig.\,\ref{fig.stiff_populations}b it can be seen that, as in the closed case at $t=0$ the population is in the trans-$S_{1}$ state, the population dynamics in the first $100$ fs proceeds in a similar manner to the closed dynamics. Population transfer at $100$ fs has a notable difference in that a small amount of population does not overcome the first barrier, located at approximately $\theta=0.3\pi$ in Fig.\,\ref{fig:PES}. This results in a small amount of population maintained in the trans-$S_{1}$ state. The majority of the population, is transferred to the cis-$S_{1}$ state with some transference, approximately $20\%$, occurring to the desired cis-$S_{0}$ state. There are some notable changes in the ensuing dynamics for the damped case. For example, following the cis-$S_{1}$ population, between $200$-$400$ fs there is a decaying transference between cis-$S_{1}$ and trans-$S_{1}$, accompanied by a small rise in trans-$S_{0}$. This is explained by the feature of the second barrier located at approximately $\theta = 0.7\pi$ in Fig.\,\ref{fig:PES}, which causes two dynamical effects. The first is that the wavepacket approaching the barrier from the perpendicular confirmation does not pass over it, and thus there is some transference back to trans-$S_{1}$ and a small amount of transference to trans-$S_{0}$. The second effect is that the wavepacket overcomes the barrier but then becomes partially trapped in the region $0.7\pi \leq \theta \leq 1\pi$. Therefore, in contrast to the closed dynamics, at $400$ fs there is a greater cis-$S_{1}$ population than trans-$S_{1}$. In addition, the population transfer to cis-$S_{0}$ at $100$ fs remains trapped over the time $100$-$400$ fs. This is due to the damped dynamics of the wavepacket on the ground PES in the region $0.5\pi \leq \theta \leq 1\pi$, the wavepacket is no longer able to reach the vicinity of the crossing point and instead relaxes to the minima of the potential at $\theta = \pi$. To summarise, the damped dynamics causes a larger cis population on both excited and ground PES at $400$ fs.

\begin{figure}
	\includegraphics{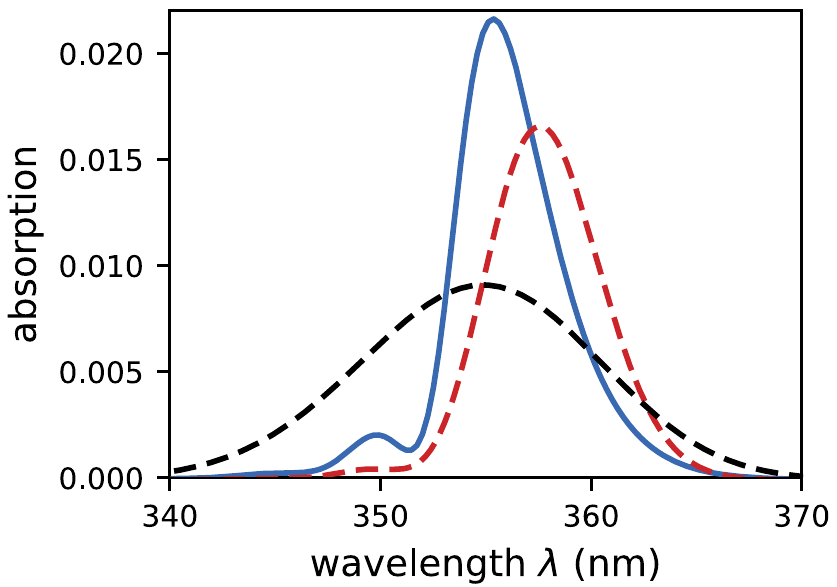}% Here is how to import EPS art
	\caption{Linear absorption spectra of stiff-stilbene using a model PES (blue line). To analyse the features of the spectra, the standard displaced harmonic model spectra (black dashed line) and differing curvature model spectra (red dashed line) are plotted. Notably, there is the appearance of the s-progression, described in Sec.\,\ref{sec:level3a}, at $\lambda = 350$ nm.  }\label{fig.stiff_spectra}
\end{figure}

The features of the PES of Fig.\,\ref{fig:PES} can also be assessed by analysing linear absorption spectra. To generate the absorption spectra, the wavepacket dynamics are first simulated and the dephasing function of Eq.\,(\ref{eq.dephasing_function}) is calculated, which is then substituted into Eq.\,(\ref{eq.absorption}), in which parallel dipoles are assumed and $\vert \mu_{eg}\vert^{2} = 1$. The results are presented in Fig.\,\ref{fig.stiff_spectra}. For comparison, absorption spectra for the standard displaced harmonic oscillator model is plotted as the black dashed line in Fig.\,\ref{fig.stiff_spectra}, assuming that $\omega_{g} = \omega_{e} = 0.01571$ fs$^{-1}$. Additionally, the absorption spectra for the differing curvature model is plotted as the red dashed line, with $\omega_{g} = 0.085 $ fs$^{-1}$ and $\omega_{e} = 0.01571$ fs$^{-1}$. The spectra of stiff-stilbene, generated using dynamics on the model PES of Fig.\,\ref{fig:PES}, is represented by the blue line in Fig.\,\ref{fig.stiff_spectra}. The stiff-stilbene spectra presents some features which are not captured by the standard displaced harmonic oscillator model. Firstly, the width of the peak is much narrower and the peak maximum is shifted to larger wavelength. Both of these features are a result of different curvature in ground and excited PES, and are present in the spectra of the differing curvature model. Secondly, there is the appearance of the s-progression described in Sec.\,\ref{sec:level3a}, at $\lambda = 350$ nm, which is also a result of different curvature and present in the differing curvature model spectra. The results presented in this section do not give rise to the well resolved vibronic progressions of Sec.\,\ref{sec:level3a} due to the presence of the environment which causes spectral broadening. In the case of the differing curvature model spectra, this broadening can make it difficult to observe the s-progression. Additionally, the s-progression feature is diminished further due to the asymmetric broadening discussed in Sec.\,\ref{sec:level3b}, which results in a larger amount of broadening for smaller wavelength. We also note that, as harmonic raising and lowering operators are used, the extent of this asymmetric broadening may be different than if the raising and lowering operators of the system manifold were used. This is dependent on the spacing of the eigenenergies, and if the anharmonicity makes the spacing smaller or larger.

Although the differing curvature model allows some explanation for the rise of features of the stiff-stilbene spectra, it does not completely capture all spectral features. For example, the stiff-stilbene spectra is less shifted towards larger wavelength. This is a result of the anharmonicity of the excited state PES, whereby the potential energy barriers, cause a shift of the eigenenergies above the barrier to larger energies \cite{Pupasov-Maksimov2016}. This, in turn, causes the spectra to be shifted to smaller wavelengths.  Furthermore, the s-progression is enhanced for the stiff-stilbene spectra. This is due to the widening of the excited state PES in the model before it rises steeply to act as a confining well. The eigenfunctions thus become elongated creating a greater overlap with higher lying states than in the differing curvature model. It should be noted that the experimental absorption spectra for stiff-stilbene exhibits more complexity due to the presence of other modes which are not directly involved in the isomerisation pathway. As a consequence, the additional peaks can also obscure spectral features for the presented band generated by torsion about the carbon double bond in stiff-stilbene. Lastly, the s-progression corresponds to energies above the initial point of excitation, at $S_{1}(0) = 3.5$ eV, which can be associated with wavelengths of $\lambda \leq 354$ nm. Therefore, it is possible that a continuum of states contributes to the spectra in this region. 
\section{\label{sec:level4}Conclusion}
We presented a model PES for stiff-stilbene in Fig.\,\ref{fig:PES}, inspired by a schematic diagram and TD-DFT data \cite{Quick2014,Improta2005}. In addition to this, we incorporated the effects of an environment through a stochastic Schr\"{o}dinger equation approach. Subsequently, two prominent features of the PES were identified. The first was the large difference in curvature of the excited and ground PES accompanied by a large displacement, the second feature was anharmonicity in the form of potential energy barriers. The first feature was studied in Sec.\,\ref{sec:level3a}, using a harmonic oscillator with differing curvatures model. This revealed the presence of a s-progression, a substructure in the vibronic progression, which was subsequently explained and quantified via a derived expression for the Franck-Condon coefficients. Linear absorption spectra of an anharmonic dissipative system was then studied using the Morse potential in Sec.\,\ref{sec:level3b}. This revealed the sensitivity of spectral features due to the combined effects of asymmetric broadening, and alteration of vibronic progression intensity and spacing caused by anharmonicity. Furthermore, the assumption of harmonic raising and lowering operators used as the Lindblad operators that define interaction with the environment was tested using analytic expressions. This provided the observation that using harmonic raising and lowering operators causes a greater broadening for higher frequencies than using the Morse counterpart raising and lowering operators, though such a feature was only prominent for large displacements. Lastly, the population dynamics, and absorption spectra, generated using the model PES for stiff-stilbene was analysed. The former displayed the importance of an interplay of anharmonicity in the form of potential energy barriers, and damped dynamics. This suggests a photoselectability of cis and trans states that is dependent on a tuning of interaction with the environment and anharmonicity. The absorption spectra presented spectral features, of the model stiff-stilbene PES, in the form of the s-progression, a decrease of peak width, and shift of peak maximum to larger wavelengths. These features were largely accounted for by the difference in curvature in ground and excited PES and the large displacement between potentials. However, the presence of the potential energy barriers additionally caused the spectra to be less shifted to larger wavelength than in the differing curvature model.

The presence of the s-progression in experimental absorption spectra presents a method of verifying, or estimating, the difference in curvature of ground and excited PES. This would be achieved by measuring the width of the s-progression, then comparing and fitting to the derived expressions for the FC coefficients.

The results here demonstrate population dynamics and spectral features present in a model PES for stiff-stilbene. Follow-up research may consider a more realistic treatment of stiff-stilbene that, for example, accounts for other modes and a continuum of states. This would give rise to more complexity in the absorption spectra, which may obscure or diminish the s-progression.  Additionally, temperature effects and solvent properties that more closely align with experiment may be included. This would allow for a rigorous assessment of the spectral features of stiff-stilbene and analysis of the importance of potential energy barriers and environment effects.  

\section*{Acknowledgements}
LDS received funding from the Engineering and Physical Sciences Research Council.

\appendix
\section{\label{sec:level5a} stiff-stilbene model PES}
We begin by describing the additional terms of the ground potential of Fig.\,\ref{fig:PES}. The confining well term of Eq.\,(\ref{eq.S0}) is given by
\begin{equation}
S_{01}(\theta) = \lambda_{g} \sin(\theta) + \mu_{g}\bigg(\frac{1-\cos(1/2(\theta - \pi/2))}{1-(1/\sqrt{2})}-1\bigg),
\end{equation}
where we choose $\lambda_{g}=12(1-1/\sqrt{2})$ and $\mu_{g} = 10(1-1/\sqrt{2})$, which control the steepness of the well and also the position of the minima. This confining well ensures that the wavepacket is confined to the regions of Fig.\,\ref{fig:PES}. It is possible to reduce the number of equations by setting $\lambda_{g} = \mu_{g} + E_{P}$. In which case, Eq.\,(\ref{eq.S0}) becomes redundant and the ground potential is described by
\begin{equation}
S_{01}(\theta) = (\mu_{g} + E_{P}) \sin(\theta) + \mu_{g}\bigg(\frac{1-\cos(1/2(\theta - \pi/2))}{1-(1/\sqrt{2})}-1\bigg).
\end{equation} 
If the minima of this potential placed at the points $\theta = 0$ and $\theta = \pi$ is desired, differentiation provides further reduction of parameters to 
\begin{equation}
\mu_{g} = \frac{E_{P}}{2}(\sqrt{2}-1).
\end{equation}
We make use of the unreduced form Eq.\,(\ref{eq.S0}) to allow for a potential that is of the form $1-\cos(2\theta)$ with an additional confining well term. This allows control over the energy at the perpendicular conformation, whilst also allowing control over the symmetry about the minima of the wells at $\theta = 0$ and $\theta = \pi$.\

We now describe the additional terms $S_{11}(\theta)$, $S_{12}(\theta)$, and $S_{13}(\theta)$ of the excited potential $S_{1}(\theta)$, which is displayed in Fig.\,\ref{fig:PES}. Starting with Eq.\,(\ref{eq.S1}), the cosine amplitude is chosen as $\eta_{e}=0.0702$. The first additional term, which describes the confining well, is given by 
\begin{equation}
S_{11}(\theta) = \lambda_{e} \sin(\theta) + \mu_{e}\bigg(\frac{1-\cos(1/2(\theta - \pi/2))}{1-(1/\sqrt{2})}-1\bigg),
\end{equation}
where we choose $\lambda_{e}=17(1-1/\sqrt{2})$ and $\mu_{e}=15(1-1/\sqrt{2})$. The second term, which allows control over the well depth in the region of interest, is described by
\begin{equation}
S_{12}(\theta) = -\xi_{e}(1-\cos(2\theta)),
\end{equation}
which is akin to an inverted form of the ground PES, where $\xi_{e}=0.375$ controls the well depth.  The final term, is given by
\begin{equation}
S_{13}(\theta) = \zeta_{e}\sin(4\theta),
\end{equation}
which raises the barrier in the cis conformation, whilst lowering the barrier height in the trans conformation, where $\zeta_{e}=0.00807$ controls the barrier height difference. 
\section{\label{sec:level5b} Derivation of FC coefficients}
For the derivation of FC coefficients  in the case of $D=0$ and even $n$ the explicit representation of the Hermite polynomial for even $n$ \cite{Weisstein}
\begin{equation}
H_{n}(x) = n! \sum_{l=0}^{\frac{n}{2}} \frac{(-1)^{\frac{n}{2}-l}}{(2l)!(\frac{n}{2}-l)!}(2x)^{2l}
\end{equation}
is required. Substituting this expression into Eq.\,(\ref{eq.unsolved_eq}) gives
\begin{align}
\langle \psi_{g}^{n=0} \vert  \psi_{e}^{n}\rangle=& N_{e}N_{n} n! \sum_{l=0}^{\frac{n}{2}} \frac{(-1)^{\frac{n}{2}-l}}{(2l)!(\frac{n}{2}-l)!} \nonumber\\ &\times\int_{-\infty}^{\infty}dx\,(2\sqrt{\alpha_{e}}x)^{2l}\exp\bigg(-\alpha x^{2}\bigg) \label{eq.gaussian_sub}.
\end{align}
The integral in Eq.\,(\ref{eq.gaussian_sub}) is a gaussian integral with a known result
\begin{equation}
\int_{-\infty}^{\infty}dx\,x^{2n}\exp\bigg(-a x^{2}\bigg) = \frac{(2n-1)!!}{2^{n}a^{n}}\sqrt{\frac{\pi}{a}},
\end{equation}
where $!!$ represents the double factorial, for which by definition $(-1)!! = 0!! = 1$. Substituting the solution to the integral gives
\begin{align}
\langle \psi_{g}^{n=0} \vert  \psi_{e}^{n}\rangle=& N_{e}N_{n} n! \sum_{l=0}^{\frac{n}{2}} \frac{(-1)^{\frac{n}{2}-l}}{(2l)!(\frac{n}{2}-l)!} \nonumber\\ &\times (2\sqrt{\alpha_{e}})^{2l}\frac{(2l-1)!!}{2^{l}\alpha^{l}}\sqrt{\frac{\pi}{\alpha}} \label{eq.double_fact_sub}
\end{align}
Further simplification can be achieved by using the following definition for the double factorial
\begin{equation}
(2n - 1)!! = \frac{(2n)!}{2^{n}n!}.
\end{equation}
Substituting this expression into Eq.\,(\ref{eq.double_fact_sub}) and simplifying gives
\begin{align}
\langle \psi_{g}^{n=0} \vert  \psi_{e}^{n}\rangle=& N_{e}N_{n} n!\sqrt{\frac{\pi}{\alpha}} \sum_{l=0}^{\frac{n}{2}} \frac{(-1)^{\frac{n}{2}-l}}{(l)!(\frac{n}{2}-l)!}\bigg(\frac{\alpha_{e}}{\alpha}\bigg)^{l}
\end{align}
By taking a factor of $(n/2)!$ out of the summation we can recast the equation in terms of a binomial coefficient 
\begin{align}
\langle \psi_{g}^{n=0} \vert  \psi_{e}^{n}\rangle=& N_{e}N_{n} \frac{n!}{(\frac{n}{2})!}\sqrt{\frac{\pi}{\alpha}} \sum_{l=0}^{\frac{n}{2}} {\frac{n}{2}\choose l}  \bigg(\frac{\alpha_{e}}{\alpha}\bigg)^{l}.
\end{align}
The binomial formula 
\begin{align}
\sum_{k=0}^{n} {n \choose k} x^{n-k} y^{k} 
\end{align}
is now used to further simplify the equation to give
\begin{align}
\langle \psi_{g}^{n=0} \vert  \psi_{e}^{n}\rangle=& N_{e}N_{n} \frac{n!}{(\frac{n}{2})!}\sqrt{\frac{\pi}{\alpha}}\bigg(\frac{\alpha_{e}}{\alpha}-1 \bigg)^{\frac{n}{2}}. \label{eq.D=0_result}
\end{align}
Substituting Eq.\,(\ref{eq.norm_g}) and Eq.\,(\ref{eq.norm_e}), simplifying, and taking the square absolute value provides the final expression for the Franck-Condon coefficient
\begin{align}
\vert \langle \psi_{g}^{n=0} \vert  \psi_{e}^{n}\rangle \vert^{2}=&  \frac{n!}{2^{n}((\frac{n}{2})!)^{2}}\frac{\sqrt{\alpha_{e}\alpha_{g}}}{\alpha}\bigg(1-\frac{\alpha_{e}}{\alpha} \bigg)^{n}.
\end{align}

For the more general case of $D\geq 0$ and $\omega_{g} \neq \omega_{e}$ Eq.\,(\ref{eq.overlap_recast}) can be recast into the form \cite{Chang2005}
\begin{align}
\langle \psi_{g}^{n=0} \vert  \psi_{e}^{n}\rangle=& N_{e}N_{n}\exp\bigg(-\frac{S}{2}\bigg) \int_{-\infty}^{\infty}dx\,H_{n}(\sqrt{\alpha_{e}}(x-d)) \nonumber \\
&\times \exp\bigg(-\frac{\alpha_{g} + \alpha_{e}}{2}\bigg(x -\frac{\alpha_{e}d}{\alpha_{g}+\alpha_{e}}\bigg)^{2}\bigg), \label{eq.y_recast}
\end{align}
where
\begin{equation}
A = \frac{\alpha_{g}\alpha_{e}d^{2}}{\alpha_{g}+\alpha_{e}}.
\end{equation}
Next let 
\begin{equation}
y = x - \frac{\alpha_{e}d}{\alpha_{g} + \alpha_{e}},
\end{equation}
and substitute this expression into Eq.\,(\ref{eq.y_recast}) to give
\begin{align}
\langle \psi_{g}^{n=0} \vert  \psi_{e}^{n}\rangle=& N_{e}N_{n}\exp\bigg(-\frac{A}{2}\bigg) \int_{-\infty}^{\infty}dy\,H_{n}(\sqrt{\alpha_{e}}y + \beta) \nonumber\\
&\times \exp\big(-\alpha y^{2}\big), \label{hermite_sub}
\end{align}
where
\begin{equation}
\alpha = \frac{\alpha_{g} + \alpha_{e}}{2}
\end{equation}
as before and 
\begin{equation}
\beta = \frac{\alpha_{g}\sqrt{\alpha_{e}}d}{\alpha_{g}+\alpha_{e}}.
\end{equation}
A Taylor expansion of the Hermite polynomial provides the useful property
\begin{align}
H_{n}(x+y) = \sum_{k=0}^{n}H_{k}(\sqrt{\alpha_{e}}y)(2\beta)^{n-k}.
\end{align}
Using this property, for the Hermite polynomial in Eq.\,(\ref{hermite_sub}), gives
\begin{align}
\langle \psi_{g}^{n=0} \vert  \psi_{e}^{n}\rangle=& N_{e}N_{n}\exp\bigg(-\frac{A}{2}\bigg)\sum_{k=0}^{n}(2\beta)^{n-k} \nonumber\\ \times&\int_{-\infty}^{\infty}dy\,H_{k}(\sqrt{\alpha_{e}}y)
\times \exp\big(-\alpha y^{2}\big),
\end{align}
where the integral in the equation is of the same form as Eq.\,(\ref{eq.unsolved_eq}). Therefore, by using the result in Eq.\,(\ref{eq.D=0_result}), of the derivation for the case of no displacement we obtain
\begin{align}
\langle \psi_{g}^{n=0} \vert  \psi_{e}^{n}\rangle=& N_{e}N_{n} \sqrt{\frac{\pi}{\alpha}}\exp\bigg(-\frac{A}{2}\bigg)\sum_{k=0}^{n}{n \choose k}(2\beta)^{n-k} \nonumber\\&\times \frac{k!}{(\frac{k}{2})!}\bigg(\frac{\alpha_{e}}{\alpha}-1\bigg)^{k/2},
\end{align}
for even $k$. For the purpose of clarity, we now replace $k$ with $2l$ for $l \in \mathbb{N}_{0}$ to give, 
\begin{align}
\langle \psi_{g}^{n=0} \vert  \psi_{e}^{n}\rangle=& N_{e}N_{n} \sqrt{\frac{\pi}{\alpha}}\exp\bigg(-\frac{A}{2}\bigg)\sum_{l=0}^{\left \lfloor{n/2}\right \rfloor }{n \choose 2l}(2\beta)^{n-2l} \nonumber \\&\times \frac{2l!}{(l)!}\bigg(\frac{\alpha_{e}}{\alpha}-1\bigg)^{l},
\end{align}
where the floor function of $n/2$ has been taken in the upper limit of the summation such that double counting does not occur, and to ensure that $2l$ represents an even number.
Using the binomial coefficient formula, rearranging and simplifying, gives the form
\begin{align}
\langle \psi_{g}^{n=0} \vert  \psi_{e}^{n}\rangle=& N_{e}N_{n} \sqrt{\frac{\pi}{\alpha}}\exp\bigg(-\frac{A}{2}\bigg) \nonumber\\ &\times n!\sum_{l=0}^{\left \lfloor{n/2}\right \rfloor }\frac{(-1)^{l}}{l!(n-2l)!}(2\beta)^{n-2l} \bigg(1-\frac{\alpha_{e}}{\alpha}\bigg)^{l},
\end{align}
the motivation for which shall become clear in imminent discussion. Firstly, we substitute for $N_{g}$ and $N_{n}$, using Eq.\,(\ref{eq.norm_g}) and Eq.\,(\ref{eq.norm_e}),
\begin{align}
\langle \psi_{g}^{n=0} \vert  \psi_{e}^{n}\rangle=& \frac{1}{\sqrt{2^{n}n!}}\bigg(\frac{\sqrt{\alpha_{e}\alpha_{g}}}{\alpha}\bigg)^{1/2}\exp\bigg(-\frac{A}{2}\bigg) \nonumber\\ &\times n!\sum_{l=0}^{\left \lfloor{n/2}\right \rfloor }\frac{(-1)^{l}}{l!(n-2l)!}(2\beta)^{n-2l} \bigg(1-\frac{\alpha_{e}}{\alpha}\bigg)^{l}.
\end{align}
Thus, the Franck-Condon coefficients for the differing curvature model, that admits displacement, is given by 
\begin{align}
\vert \langle \psi_{g}^{n=0} \vert  \psi_{e}^{n}\rangle \vert^{2}=& \frac{1}{2^{n}n!}\frac{\sqrt{\alpha_{e}\alpha_{g}}}{\alpha}e^{-A} \nonumber \\
&\times \bigg\vert n!\sum_{l=0}^{\left \lfloor{n/2}\right \rfloor }\frac{(-1)^{l}}{l!(n-2l)!}(2\beta)^{n-2l} \nonumber\\
&\times \bigg(1-\frac{\alpha_{e}}{\alpha}\bigg)^{l} \bigg\vert^{2}.
\end{align}
\bibliography{Quantum_dissipative_systems_beyond_the_standard_harmonic_model_features_of_linear_absorption_and_dynamics}
\end{document}